\documentclass[a4paper,fleqn]{cas-dc}

\usepackage[numbers]{natbib}

\usepackage{amsmath,amssymb,amsfonts}
\usepackage{mathtools}
\usepackage{bm}
\usepackage{amsthm}
\usepackage{textcomp}

\usepackage{graphicx}
\usepackage{caption}
\usepackage{subcaption}
\usepackage{booktabs}
\usepackage{threeparttable}
\usepackage{tabularx}
\usepackage{multirow}
\usepackage{longtable}
\usepackage{tablefootnote}
\usepackage{float}
\usepackage{wrapfig}
\usepackage{lscape}
\usepackage{stfloats}
\usepackage{placeins}
\usepackage{xcolor,colortbl}
\usepackage{makecell}
\usepackage{arydshln}

\usepackage{algorithm}
\usepackage{algcompatible}
\usepackage[noend]{algpseudocode}

\usepackage{enumerate}
\usepackage[shortlabels]{enumitem}
\usepackage{xspace}
\usepackage{calc}
\usepackage{fancyhdr}
\usepackage{varioref}
\usepackage{lipsum}
\usepackage{url}
\usepackage{verbatim}
\usepackage[normalem]{ulem}
\usepackage{soul}
\usepackage{times}

\usepackage{tikz}
\usetikzlibrary{patterns}
\usepackage{pgfgantt}

\usepackage{pifont}

\usepackage{hyperref}

\newcommand{\algrule}[1][.2pt]{\par\vskip.5\baselineskip\hrule height #1\par\vskip.5\baselineskip}
{\bfseries}{\rmfamily}
{\bfseries}{\rmfamily}
\newtheorem{mylemma}{Lemma}{\bfseries}{\rmfamily}
\newtheorem{definition}{Definition}

\newcommand{\qpadl}{\ensuremath {\texttt{QPADL}}}
\newcommand{\qpadlhct}{\ensuremath {\texttt{QPADL\mbox{-}HCT}}}
\newcommand{\qpadllbp}{\ensuremath {\texttt{QPADL\mbox{-}LBP}}}
\newcommand{\qpadloop}{\ensuremath {\texttt{QPADL\mbox{-}OOP}}}
\newcommand{\qpadlenc}{\ensuremath {\texttt{QPADL\mbox{-}ENS}}}
\newcommand{\qpadlftr}{\ensuremath {\texttt{QPADL\mbox{-}FTR}}}
\newcommand{\hcp}{\ensuremath {\texttt{HCT}}}
\newcommand{\lbp}{\ensuremath {\texttt{LBP}}}

\newtheorem{theorem}{Theorem}
\newcommand{\as}{\ensuremath {\leftarrow}{\xspace}}
\newcommand{\pqtor}{\texttt{PQ}\mbox{-}\texttt{Tor}}

\newcommand{\hct}{\ensuremath {\texttt{HCT}{\xspace}}}
\newcommand{\lb}{\ensuremath {\texttt{LB}{\xspace}}}
\newcommand{\puzgen}{\ensuremath {\texttt{Puzzle.Gen}{\xspace}}}
\newcommand{\pow}{\ensuremath {\texttt{PoW}{\xspace}}}
\newcommand{\powverif}{\ensuremath {\texttt{PoW.Verify}{\xspace}}}
\newcommand{\client}{\ensuremath {\texttt{Client}{\xspace}}}
\newcommand{\psd}{\ensuremath {\texttt{PSD}}}

\newcommand{\ap}{\ensuremath {\texttt{AP}}}
\newcommand{\id}{\ensuremath {\texttt{ID}}}
\newcommand{\su}{\ensuremath {\texttt{SU}}}
\newcommand{\blockrec}{\ensuremath {\texttt{BlockReconst}{\xspace}}}
\newcommand{\dbindex}{\ensuremath {\texttt{DB.Index}{\xspace}}}

\newcommand{\keygen}{\ensuremath {\texttt{KeyGen}{\xspace}}}
\newcommand{\verify}{\ensuremath {\texttt{Verify}{\xspace}}}
\newcommand{\sign}{\ensuremath {\texttt{Sign}{\xspace}}}

\newcommand{\asrandom}{\ensuremath{\stackrel{\$}{\leftarrow}}}

\newcommand{\zq}{\ensuremath \mathbb{Z}_{q}{\xspace}}

\newcommand{\sk}{\ensuremath {\mathit{sk}}{\xspace}}
\newcommand{\pk}{\ensuremath {\mathit{PK}}{\xspace}}

\newcommand{\db}{\ensuremath {\texttt{DB}}}

\newcommand{\kyber}{\ensuremath {\texttt{Kyber}{\xspace}}}
\newcommand{\mlkem}{\ensuremath {\texttt{ML-KEM}{\xspace}}}

\newcommand{\mldsa}{\ensuremath {\texttt{ML-DSA}{\xspace}}}

\newcommand{\ts}{\ensuremath {\texttt{TS}}}
\newcommand{\rtt}{\ensuremath {\texttt{RTT}}}
\newcommand{\rss}{\ensuremath {\texttt{RSS}}}
\newcommand{\envv}{\ensuremath {\texttt{env}}}
\newcommand{\tw}{\ensuremath {\texttt{TW}}}
\newcommand{\ch}{\ensuremath {\texttt{ch}}}
\newcommand{\tv}{\ensuremath {\texttt{TV}}}

\newcommand{\createtoken}{\ensuremath {\texttt{Create.Token}{\xspace}}}
\newcommand{\query}{\ensuremath {\texttt{Query}{\xspace}}}
\newcommand{\response}{\ensuremath {\texttt{Response}{\xspace}}}
\newcommand{\pir}{\ensuremath {\texttt{PIR}{\xspace}}}
\newcommand{\circuit}{\ensuremath{\texttt{Circuit.Creation}{\xspace}}}

\newcommand{\setup}{\ensuremath{\texttt{Setup}{\xspace}}}

\newcommand{\service}{\ensuremath {\texttt{Service}{\xspace}}}

\newcommand{\request}{\ensuremath {\texttt{Request}{\xspace}}}
\newcommand{\gen}{\ensuremath {\texttt{Gen}{\xspace}}}
\newcommand{\lrs}{\ensuremath {\texttt{LRS}{\xspace}}}
\newcommand{\link}{\ensuremath {\texttt{Link}{\xspace}}}
\newcommand{\pol}{\ensuremath {\texttt{PoL}{\xspace}}}
\newcommand{\ProxVerif}{\ensuremath {\texttt{ProxVerif}{\xspace}}}
\newcommand{\qvec}{\ensuremath {\boldsymbol{q}}}
\newcommand{\qlen}{\ensuremath {\boldsymbol{\Bar{q}}}}
\newcommand{\rvec}{\ensuremath {\boldsymbol{\rho}}}
\newcommand{\squarenum}[1]{%
  \tikz[baseline=(char.base)]\node[
    shape=rectangle,
    draw=blue,
    fill=black,
    text=white,
    inner sep=2pt,
    rounded corners=2pt,
    scale=0.8
  ] (char) {\sffamily\bfseries\small #1};%
}

\def\tsc#1{\csdef{#1}{\textsc{\lowercase{#1}}\xspace}}
\tsc{WGM}
\tsc{QE}

\begin{document}
\let\WriteBookmarks\relax
\def\floatpagepagefraction{1}
\def\textpagefraction{.001}

\shorttitle{QPADL: Post-Quantum Private Spectrum Access with Verified Location and DoS Resilience}    

\shortauthors{Darzi et al.}  

\title [mode = title]{QPADL: Post-Quantum Private Spectrum Access with Verified Location and DoS Resilience}



\author[1]{Saleh Darzi}
\ead{salehdarzi@usf.edu}

\author[1]{Saif Eddine Nouma}
\ead{saifeddinenouma@usf.edu}

\author[1]{Kiarash Sedghighadikolaei}
\ead{kiarashs@usf.edu}

\author[1]{Attila Altay Yavuz}
\ead{attilaayavuz@usf.edu}

\affiliation[1]{organization={Bellini College of Artificial Intelligence, Cybersecurity and Computing, University of South Florida},
            city={Tampa},
            postcode={33620},
            state={Florida},
            country={USA}}
            

\begin{abstract}
With advances in wireless communication and growing spectrum scarcity, Spectrum Access Systems (SASs) offer an opportunistic solution but face significant security challenges. Regulations require disclosure of location coordinates and transmission details, exposing user privacy and anonymity during spectrum queries, while the database operations themselves permit Denial-of-Service (DoS) attacks. As location-based services, SAS is also vulnerable to compromised or malicious users conducting spoofing attacks. These threats are further amplified given the advances in quantum computing. Thus, we propose $\qpadl$, the first post-quantum (PQ) secure framework that simultaneously ensures privacy, anonymity, location verification, and DoS resilience while maintaining efficiency for large-scale spectrum access systems. $\qpadl$ introduces SAS-tailored private information retrieval for location privacy, a PQ-variant of Tor for anonymity, and employs advanced signature constructions for location verification alongside client puzzle protocols and rate-limiting technique for DoS defense. We formally assess its security and conduct a comprehensive performance evaluation, incorporating GPU parallelization and optimization strategies to demonstrate practicality and scalability.
\end{abstract}




\maketitle

\section{Introduction} \label{sec:Intro}
The rapid growth of wireless technologies (e.g., mobile and IoT) combined with regulated spectrum allocation (e.g., FCC in the U.S.~\cite{grissa2016preserving}) has led to spectrum scarcity. Cognitive Radio Networks (CRNs) mitigate this challenge by enabling Secondary Users (SUs) to opportunistically access unused licensed channels~\cite{grissa2021anonymous}. At the core of this model lies the Spectrum Access System (SAS), which dynamically allocates spectrum through geo-location databases (DBs)~\cite{chakraborty2023capow}. Despite its effectiveness, SAS poses major privacy and security risks: continuous reporting of locations and transmission details compromises user privacy and anonymity~\cite{jasim2021cognitive}, while its location-centric design enables spoofing, falsified data, and unauthorized access~\cite{xin2016privacy}. The broadcast nature of spectrum communication, reliance on databases, and widespread use of low-cost devices further expose SAS to denial-of-service (DoS) attacks~\cite{darzi2024privacy}. These challenges are intensified by quantum computing, which threatens classical cryptographic protections~\cite{bavdekar2023post}.  \textit{While prior efforts to address these security challenges in spectrum access remain isolated, no existing work offers a unified PQ-secure framework that simultaneously ensures location privacy, anonymity, verifiable location, and DoS resilience under FCC-compliant SAS constraints}. 

\subsection{Related Works and Open Problems} \label{subsec:literature}

\noindent{\em (i) {\textbf{Location Privacy and Anonymity in SAS:}}} 
Centralized SAS, mandated by the FCC, operates through multiple geolocation databases to facilitate dynamic spectrum sharing among governmental, commercial, licensed and unlicensed users~\cite{grissa2016preserving}. Access requires disclosure of sensitive data such as exact geographic coordinates, channel preferences, usage patterns, and transmission parameters, leaving users vulnerable to identity tracing, behavioral profiling, and lifestyle inference attacks~\cite{jasim2021cognitive, agarwal2022survey}. However, existing approaches remain inadequate: (1) k-anonymity and pseudo-identifier methods \cite{zhu2019lightweight} lack provable security and provide only weak guarantees unless large anonymity sets are used, which is infeasible in large-scale SAS deployments; (2) differential privacy-based techniques \cite{ul2022differential}, while theoretically sound, significantly degrade the accuracy of spectrum availability information; and (3) Private Information Retrieval (PIR)-based schemes \cite{grissa2021anonymous, xin2016privacy} focus solely on location privacy while neglecting anonymity, assuming honest uncompromised users, and ignoring location spoofing. These gaps highlight the necessity for efficient, provably secure mechanisms that provide strong anonymity and location privacy under realistic network assumptions, without sacrificing performance or user experience~\cite{darzi2024privacy}.

\noindent{\em (ii) {\textbf{Location Verification and Spoofing Attack Resistance:}}} SAS, as a location-based platform, relies on accurate user location data for fair and efficient spectrum allocation, making it vulnerable to adversaries who impersonate users or submit falsified locations, leading to interference, disruptions, and economic loss~\cite{nguyen2019spoofing}. Existing location verification schemes often assume trusted infrastructure, honest participants, or dedicated location servers, assumptions that are unrealistic in rural or infrastructure-limited settings~\cite{xin2016privacy}. Furthermore, most approaches fail to preserve location privacy and anonymity from the verifying entities themselves. These limitations underscore the need for a practical and privacy-preserving location assurance framework capable of operating under real-world constraints while resisting diverse location-based attacks~\cite{darzi2025slap}.

\noindent{\em (iii) {\textbf{Counter-DoS and Spectrum Management for Next-Gen Network Systems:}}} The widespread deployment of low-cost IoT devices, together with SAS’s dependence on geolocation databases, has considerably heightened susceptibility to DoS attacks~\cite{jakimoski2008denial}. These attacks flood the system with illegitimate requests, hindering spectrum coordination and impairing overall system responsiveness, especially during spectrum access and service interactions. 

Various mitigation strategies have been proposed, including intrusion detection systems (IDSs), blockchain-based access control, and cryptographic methods such as Client Puzzle Protocols (CPPs)~\cite{darzi2024counter}. Machine learning has further advanced IDSs, enabling detection of abnormal traffic patterns with reported accuracies above 95\%~\cite{chakraborty2023capow}. However, these methods focus on detection rather than prevention, and their effectiveness depends on continuous access to sensitive user traffic and often private network topology, requirements that are impractical for real-time SAS defense. In addition, AI-driven defenses remain susceptible to adversarial manipulation, where attackers exploit model weaknesses to evade detection, potentially incurring significant operational costs~\cite{doriguzzi2024flad}.  
CPPs mitigate malicious requests by requiring clients to solve puzzles before service~\cite{bostanov2021client}, but at scale, they suffer from bottlenecks in puzzle generation, distribution, and parallelization, imposing high costs on servers and legitimate users~\cite{ali2020foundations}. Outsourcing puzzle generation, as in \cite{darzi2024privacy}, reduces server load but shifts DoS risks to spectrum databases, which attackers can still overwhelm with queries. Thus, a lightweight rate-limiting mechanism embedded in the spectrum query phase is crucial to maintain availability, counter large-scale DoS, and improve SAS resilience without excessive overhead.

\noindent{\em (iv) {\textbf{Spectrum Access in the Post-Quantum Era:}}} 
The rise of quantum computing threatens the long-term security of wireless networks by breaking classical cryptographic primitives that protect SAS~\cite{bavdekar2023post}. Protocols that preserve privacy, anonymity, and location verification rely on hardness assumptions vulnerable to quantum attacks, leaving systems exposed to threats such as location disclosure, spoofing, and large-scale DoS~\cite{darzi2024privacy}. Ensuring durable protection requires adopting Post-Quantum Cryptography (PQC) to preserve privacy, anonymity, and availability against quantum-capable adversaries.

\subsection{Our Contributions}
\label{subsec:contribution} 
{\em To the best of our knowledge, $\qpadl$ is the first PQ-secure framework for spectrum access that simultaneously achieves several often conflicting objectives, ensuring SUs' privacy and anonymity while complying with FCC's strict regulations, even under realistic network settings where malicious or compromised users may attempt location spoofing to gain spectrum advantages or launch DoS attacks to disrupt legitimate access. $\qpadl$ provides the following desirable properties:}

\smallskip
\noindent$\bullet$ {\em \ul{\textbf{Location Privacy-Preserving and Anonymous Spectrum Access}}:}  
$\qpadl$ achieves PQ-secure location privacy and anonymity of SUs in spectrum access while abiding by the FCC's regulations. We propose an NIST-compliant PQ-variant of The Onion Routing (Tor) network to anonymize spectrum queries to Private Spectrum Databases (PSDs) and design three $\qpadl$ instantiations: $\qpadlenc$ for efficiency, $\qpadlftr$ for robustness, and $\qpadloop$ for reduced communication. These instantiations are built on various information-theoretic (IT) or PQ multi-server PIR techniques and are tailored for FCC-compliant SAS.

\noindent$\bullet$ {\em \ul{\textbf{Proximity-backed Location Assurance and Spoofing Resistance}}:} $\qpadl$ operates under realistic network assumptions with malicious or compromised SUs, leveraging already existing WiFi Access Points (APs) and cellular towers during the spectrum query phase to verify SU locations and issue time-sensitive, quantum-safe proofs using signal strength for proximity checks. To our knowledge, this work is the first to employ event-oriented linkable ring signatures (LRSs) for location proofs in wireless networks, eliminating the need for trusted or dedicated infrastructure, while the event-ID’s linkability enables limiting the number of proofs acquired. $\qpadl$ protects SAS against spoofing attacks such as mafia fraud and distance hijacking~\cite{nguyen2019spoofing}, while preserving the anonymity of both SUs and APs against SAS and PSD servers.

\noindent$\bullet$ {\em \ul{\textbf{Outsourced DoS Mitigation with SAS Architecture Compliance}}:}
To overcome the limitations of existing DoS defenses and the added risks of quantum-capable adversaries, $\qpadl$ introduces an outsourced counter-DoS service built on diverse CPP protocols with PQ-secure puzzle generation (e.g., hash-based, lattice-based) handled by the already existing entities, i.e., PSDs. This design mitigates malicious SU and external adversary attacks during service requests and SAS server communications. While outsourcing shifts the DoS target to PSDs, $\qpadl$ is the first to reinforce their protection through a rate-limiting mechanism based on the linkability of the LRS, ensuring comprehensive DoS resistance with regulations compliance.

\noindent$\bullet$ {\em \ul{\textbf{Enabling Scalability Through Parallelization and Optimization}}:}  
We conducted comprehensive analytical and empirical evaluations of all $\qpadl$ instantiations, demonstrating their efficiency and scalability. By leveraging GPU parallelization, acceleration methods, and database compression, we mitigated PIR bottlenecks and reduced PSD-side costs affected by database size. With one to $2^{10}$ users, GPU acceleration in $\qpadlenc$ yields an average $2.77\times$ speedup on the PSD side and $4.18\times$ in end-to-end delay; $\qpadlftr$ achieves $4.88\times$ and $11.49\times$, respectively; and $\qpadloop$ delivers $1.71\times$ and $1.68\times$. These results confirm $\qpadl$’s strong scalability, particularly beyond $2^{10}$ SUs per time window.

\subsection{Improvements Over The PACDoSQ Paper\cite{darzi2024privacy}} \label{subsec:improvements} 
This work is an extended version of \textit{IEEE MILCOM-2024 Conference}~\cite{darzi2024privacy}, with $100\%$ more new material and the following significant improvements and features: 

\noindent (i)~\textbf{\textit{Location Verification Mechanism:}} We propose a PQ-secure location verification mechanism that uses LRSs and nearby WiFi APs to resist spoofing, in realistic networks with malicious or compromised SUs.

\noindent (ii)~\textbf{\textit{Comprehensive DoS Countermeasures:}} We present two $\qpadl$ instantiations using distinct CPPs and a rate-limiting technique based on LRS linkability to enhance DoS resistance against $\psd$s. The first, $\qpadlhct$, employs a hash-cash tree for tunable puzzle difficulty with parallelization resistance, while the second, $\qpadllbp$, integrates lattice-based PoWs to overcome hash-function vulnerabilities to Grover’s algorithm, offering a quantum security advantage for counter-DoS.

\noindent (iii)~\textbf{\textit{Various PIR Instantiations:}} Given PIR’s critical role in our framework, we design three $\qpadl$ instantiations: $\qpadlenc$ for efficiency, $\qpadlftr$ for fault-tolerant robustness, and $\qpadloop$ for scalability with lower communications costs.

\noindent (iv)~\textbf{\textit{Hardware Acceleration and Optimizations}} We addressed the main bottleneck of \cite{darzi2024privacy} (i.e., PSD-side computations), by applying GPU parallelization and acceleration strategies to reduce overhead and improve scalability in large-scale SAS. We also defined the DB structure, detailed setup subroutines, and proposed compression techniques.

\section{\textbf{Preliminaries and Building Blocks}}
\label{sec:Prelim} \vspace{-1mm}


%
%
\looseness-1 
\textbf{Notations:} $||$, $|x|$, and $\{0,1\}^k$ denote concatenation, the bit length of a variable, and a $k$-bit binary value, respectively. $\mathbb{F}$, $GF(2)$, and $\mathbb{Z}$ represent a finite field, the Galois Field of order 2, and the set of integers, respectively. $\{{x_i}\}_{i=1}^{\ell}$ denotes the tuple $(x_1, x_2, \dots, x_\ell)$. The notation ${x \xleftarrow{\$}\mathcal{S}}$ indicates uniform random selection from the set $\mathcal{S}$, and $\mathbf{v}$ denotes a vector. $||.||$ represents the Euclidean norm of a lattice vector. $\lambda$ is the security parameter, and $p$ is a large prime. $\Gamma$ is the Gamma function in the lattice distribution, while $\Lambda$ denotes the lattice with dimension $n_\Lambda$. The functions $H$ and $H'$ are cryptographically secure hash functions. $Enc_\sk(m)$ denotes encryption of a message $m$ using the secret key $\sk$, while $sk$ and $pk$ denote the secret and public keys, respectively. $\sigma$, $ctxt$, and $\id$ refer to the signature, ciphertext, and user identity, respectively. 

\subsection{\textbf{System Architecture}} \label{subsec:systemmodel} 
\noindent Our system model consists of the following main entities within a regulated SAS-oriented spectrum-sharing framework, where unlicensed users request transmission authorization from certified geolocation databases based on their operational parameters~\cite{PATIL2024110697}. This setting is consistent with practical deployments such as U.S. TV White Space (TVWS) and Citizens Broadband Radio Service (CBRS) in the 3.5 GHz band, in which spectrum-access decisions are mediated by authorized database administrators under regulatory oversight~\cite{caleffi2014database}. \vspace{-2mm}

\begin{itemize}[leftmargin=*]
    \item \textbf{Spectrum Regulator:} The spectrum regulator (e.g., FCC) is the governing authority of the SAS. It defines and enforces operational rules, including incumbent protection, registration and reporting requirements, and transmission constraints, while certifying and auditing database operators and providing the regulatory parameters needed for spectrum-access decisions.\vspace{-2mm}

    \item \textbf{Servers:} Servers denote application- or cloud-layer service providers (e.g., CRN platforms, web services, edge or cloud servers) that users access after obtaining spectrum authorization. They provide network services but are not involved in spectrum query and allocation. \vspace{-2mm} 

    \item \textbf{Private Spectrum Databases (PSDs):} PSDs are regulator-certified third-party geolocation DB operators (e.g., Google, Spectrum Bridge) that maintain incumbent records, protection data, and spectrum-access policies~\cite{agarwal2022survey,chen2015protocol}. Acting as the core decision entities in the SAS, they process client queries against regulatory and spectrum-usage data and return channel availability, transmission parameters, and validity intervals, while synchronizing regularly to preserve consistency under regulatory requirements~\cite{grissa2021anonymous}.\vspace{-2mm}

    \item \textbf{Clients:} Clients are unlicensed wireless devices (e.g., mobile phones, laptops, IoT nodes, or fixed terminals) that seek network access by querying PSDs for spectrum availability. After receiving authorization, they may opportunistically use licensed spectrum under the assigned operating constraints so as to avoid harmful interference to primary users. In this paper, the terms \emph{client}, \emph{user}, and \emph{SU} are used interchangeably. \vspace{-2mm}

    \item \textbf{Access Points (APs):} $\ap$ denotes the set of wireless infrastructure nodes in the region, such as WiFi access points, LTE eNodeBs, and 5G gNodeBs, that provide last-hop connectivity for clients. These APs enable communication with PSDs and servers through the network backhaul and, in our setting, may additionally support synchronized operation for proximity-backed location assurance while remaining separate from spectrum decisions~\cite{agarwal2022survey}. 
\end{itemize}

%
%

\subsection{\textbf{Cryptographic Building Blocks}} \label{subsec:threatmodel} 


\textbf{(1) Private Information Retrieval:}  
PIR enables a client to retrieve a data block from a database $\db$ without revealing its index~\cite{chor1998private}. In our architecture with synchronized spectrum DBs, we employ a multi-server PIR model where non-colluding servers each hold a copy of the DB~\cite{goldberg2007improving}.

\begin{definition} \label{def:PIR}
A multi-server PIR involves three algorithms executed between a client and $n$ non-colluding servers: \vspace{-2mm}
\begin{itemize}[leftmargin=*]
    \item[-] $\{q_i\}_{i=1}^{n} \as \client.\query(\theta, n)$: The client generates $n$ queries for index $\theta$ and sends each to database server $\db_i$. \vspace{-5mm}
    
    \item[-] $\rho_i \as \db.\query.\response(q_i)$: Each database server processes its query and returns a response $\rho_i$ derived from its local database. \vspace{-2mm}

    \item[-] $D_\theta \as \client.\blockrec(\{\rho_i\}_{i=1}^{\ell})$: The client reconstructs $D_\theta$ from the $\ell$ responses received from the database servers.

\end{itemize}
\end{definition}

\textbf{(2) Proof of Work:} 
\noindent A PoW is a cryptographic mechanism based on puzzles with tunable difficulty~\cite{back2002hashcash} that are computationally easy to generate but hard to solve. 

\begin{definition} \label{def:PoW} 
A PoW scheme operate as follows: \vspace{-2mm} 
\begin{itemize}[leftmargin=*]
    \item[-] ${\Pi \as \puzgen(1^\lambda, \kappa)}$: Given the security parameter $\lambda$ and difficulty level $\kappa$, the algorithm generates a puzzle instance $\Pi$. \vspace{-2mm} 
    \item[-] $\Psi \as \pow(\Pi, \kappa)$: Given a puzzle $\Pi$ and difficulty level $\kappa$, this algorithm computes a valid solution $\Psi$. \vspace{-2mm}
    \item[-] $\{0,1\} \as \powverif(\Pi, \Psi)$: The algorithm returns $1$ if $\Psi$ is a valid solution to puzzle $\Pi$, and $0$ otherwise.
\end{itemize}
\end{definition}

\textbf{(3) PQ-Secure Signatures:}  
To address quantum threats, NIST standardized three digital signatures under its PQC initiative~\cite{bavdekar2023post}: ML-DSA (FIPS 204~\cite{dang2024module}), FN-DSA~\cite{fouque2018falcon}, and SLH-DSA (FIPS 205~\cite{cooper2024stateless}). In $\qpadl$, we adopt ML-DSA (Module LWE/SIS), with three algorithms: $(\sk, \pk)\as \mldsa.\keygen(1^\lambda)$, $\sigma\as \mldsa.\sign(\sk, m)$, and $\{0,1\} \as \mldsa.\verify$ $(\pk, m, \sigma)$.

Ring signatures allow a user to anonymously sign a message on behalf of a group, while linkable ring signatures (LRS) further enable the detection of multiple signatures generated by the same signer within a given context. In event-oriented LRS, an event identifier is embedded in each signature, enabling linkability across signatures produced with the same key and event identifier, without revealing the signer’s identity~\cite{xue2024efficient}. 

\begin{definition} \label{def:LRS} 
An $\lrs$ scheme consists of five algorithms: \vspace{-2mm}
\begin{itemize}[leftmargin=*]
    \item[-] ${pp \as \lrs.\gen(1^\lambda)}$: Given the security parameter $\lambda$, it outputs the public parameters $pp$. \vspace{-2mm}
    \item[-] ${(\sk,\pk) \as \lrs.\keygen(pp)}$: On input $pp$, it returns a pair of private and public keys ($\sk, \pk$). \vspace{-2mm}
    \item[-] ${\sigma \as \lrs.\sign(e_\id, \sk_{l_r}, m, \mathsf{R})}$: Given event identifier $e_\id$, ring member’s secret key $\sk_{l_r}$, message $m$, and public key list $\mathsf{R}$, it outputs a signature $\sigma$. \vspace{-2mm}
    \item[-] ${\{0,1\} \as \lrs.\verify(e_\id, \sigma, m, \mathsf{R})}$: It takes $e_\id$, a signature $\sigma$ on the message $m$, and a list of public keys $\mathsf{R}$, and outputs $1$ or $0$ representing accept and reject.  \vspace{-2mm}
    \item[-] ${\{0,1\} \as \lrs.\link(e_\id, \sigma, \sigma', m, m', \mathsf{R}, \mathsf{R}')}$: Given $e_\id$, signatures ($\sigma, \sigma')$ on messages ($m, m'$), and public key lists ($\mathsf{R}, \mathsf{R}'$), it returns $1$ if linked, $0$ otherwise.
\end{itemize}
\end{definition}

We adopt a PQ secure event-oriented LRS scheme \cite{xue2024efficient} built from a hash function and Signature of Knowledge (SoK) primitives (e.g., STARK-based SoK (ethSTARK)~\cite{team2021ethstark}) with offline/online signing and verification. The signer proves in zero knowledge that (i) their public key is a leaf in a Merkle tree over the ring, and (ii) the tag is correctly computed from the secret key and event ID. Verification checks the event ID, Merkle root, and tag consistency. Linkability is enforced by matching tags across signatures sharing the same event ID. 

\subsubsection{PQ-variant of The Onion Routing ($\pqtor$)} \label{subsubsec:pqtor}
With the advent of NIST-standardized PQC, the PQ variant of the Tor network ($\pqtor$) retains the core architecture of conventional Tor while replacing vulnerable cryptographic components. Specifically, $AES128$ is upgraded to $AES256$ to mitigate Grover's algorithm~\cite{bonnetain2019quantum}; RSA signatures used in consensus are replaced with $\mldsa$ \cite{dang2024module}; RSA-based KEMs for circuit creation are substituted with $\mlkem$ \cite{mlkem}; and all key types (short-, medium-, and long-term) are updated to their PQ-secure equivalents. These substitutions preserve the functional design of Tor while achieving quantum resilience. 

Communication in $\pqtor$ proceeds as follows: {\em (i)} ${\pqtor.\setup}$: The client connects to a Directory Authority (DA), retrieves the latest network state, and selects the relay path in reverse: exit node $N_x$, middle node $N_m$, entry node $N_e$. {\em (ii)} ${\pqtor.\circuit}$: The client sends $\textit{CREATE}$ and $\textit{EXTEND}$ commands to establish the circuit, generating three AES keys and distributing them using $\mlkem$ encapsulation. {\em (iii)} ${\pqtor.\texttt{Send}(N_r; m)}$ transmits message $m$ to receiver $N_r$ via layered symmetric encryption, while ${\pqtor.\texttt{Receive}(N_s; m)}$ receives message $m$ from sender $N_s$. Each relay decrypts a layer and forwards the message, with the exit node ultimately delivering it to the destination.


%
%
\section{\textbf{Threat and Security Models}} \label{sec:securitymodel}

This section presents the adversarial and threat models of $\qpadl$, specifying the attacker’s capabilities, covered attack vectors, security assumptions, and system scope.

\subsection{Threat Model} \label{subsubsec:threatmodel} 

We consider a quantum-capable probabilistic polynomial-time (QPT) adversary in the $\qpadl$ system model. The adversary has full control over the wireless channel between clients and infrastructure entities, including eavesdropping, message injection, modification, replay, delay, and blocking. Malicious clients may also deviate arbitrarily from the protocol by issuing adaptive or strategically crafted queries, reusing protocol artifacts, or submitting manipulated location-related information to obtain unauthorized spectrum access or service. We assume the spectrum regulator (e.g., FCC) is trusted as the root authority for system initialization and policy definition. Access points are certified infrastructure entities with valid long-term credentials and synchronized operation for generating proof-related artifacts. Unless otherwise stated, $\psd$s and SAS servers are modeled as honest-but-curious: they correctly execute the protocol but may attempt to infer sensitive client information, such as identity, precise location, or cross-session linkability, from protocol transcripts, metadata, or repeated interactions. Under this model, we consider the following representative attack classes: \vspace{-2mm}

\begin{itemize}[leftmargin=*]
    \item \textit{\underline{Client Privacy and Anonymity Attacks}:} Adversaries, including $\psd$s, SAS servers, and external observers, may attempt to infer a client’s identity, location, or device-related attributes, or to link multiple sessions, by analyzing exchanged messages, metadata, timing patterns, or repeated protocol interactions. These attacks may involve passive observation, active probing, and correlation across the spectrum-query and service-access phases.\vspace{-2mm}

    \item \textit{\underline{Location Spoofing Attacks}:} Malicious clients may submit falsified coordinates or manipulated proximity-related information to gain unauthorized access to spectrum channels or SAS services outside their legitimate region. Relevant attacks include fake location claims, replay of prior proof artifacts, relay-assisted manipulation, distance fraud, mafia fraud, distance hijacking, and timestamp manipulation~\cite{nguyen2019spoofing,li2015privacy}. \vspace{-2mm}

    \item \textit{\underline{Denial-of-Service Attacks}:} Malicious clients or external adversaries may attempt to degrade the availability of $\psd$s or SAS servers by flooding them with bogus, replayed, or computationally expensive requests during the spectrum-query or service-request phases. Such attacks aim to exhaust computational or communication resources, delay responses, or deny timely spectrum access to legitimate users.
\end{itemize}

\subsection{Security Model} 

Given the system and threat models, \qpadl~aims to provide the following security objectives:

\begin{definition}[$t$-Private PIR] \label{def:tprivate}
A multi-server PIR with $\db$ and security parameter $\lambda$ is \emph{$t$-private} if for every ${0 < t < \ell}$ and every adversary $\mathcal{A}$ corrupting any subset ${S \subset [\ell]}$ with ${|S| \le t}$, there exists a simulator $\mathcal{S}$ such that for all query indices ${\{i_0,i_1\} \in \db}$, 
$|
\Pr[ \mathcal{A}(\mathsf{View}_S^{\texttt{PIR}}(i_0)) = 1 ]
-
\Pr[ \mathcal{A}(\mathsf{View}_S^{\texttt{PIR}}(i_1)) = 1 ]
|
\le \mathsf{negl}(\lambda),$
where $\mathsf{View}_S^{\texttt{PIR}}(i)$ denotes the joint view of the corrupted servers in $S$ when the honest client queries index $i$.
\end{definition}\vspace{-1mm}

\vspace{-2mm}
\begin{definition}[$\nu$-Byzantine-Robust PIR] \label{def:byzantine}
A PIR is $\nu$-Byzantine-robust if, for any set 
$B \subset [\ell]$ of at most $\nu$ Byzantine servers and any index $i \in \db$, $\Pr[ \blockrec_{\texttt{PIR}}(\rho_{[\,\ell\,]}(i,B)) = \mathsf{\db}[i] ] = 1$, where $\rho_{[\,\ell\,]}(i,B)$ are the answers from all $\ell$ servers with those in $B$ possibly adversarial.
\end{definition}\vspace{-1mm}

\vspace{-2.5mm}
\begin{definition}[$k$-out-of-$\ell$ PIR] \label{def:koutofell}
A PIR scheme is $k$-out-of-$\ell$ correct if, for any index $\theta \in \db$ and any subset $S \subseteq [\ell]$ with ${|S| \ge k}$, $\Pr[ \blockrec_{\texttt{PIR}}(\rho_S(\theta)) = \db[\theta] ] = 1$, where $\rho_S(\theta)$ are the answers from servers in $S$.
\end{definition}\vspace{-1mm}

\vspace{-2.5mm}
\begin{definition}[$\pqtor$ Anonymity] \label{def:anonymity} \looseness-1 
A $\pqtor$ network provides anonymity if any quantum-capable adversary cannot distinguish, beyond negligible probability in $\lambda$, between two executions differing only in the honest sender (or receiver), assuming all layers use PQ-secure KEMs and symmetric ciphers.
\end{definition}\vspace{-1mm}

\vspace{-2.5mm}
\begin{definition}[Correctness and Soundness of Location Verification] \label{def:locationverification}
A location verification scheme is \emph{correct} if, for any honest user at $(l_x,l_y)$, the proof of location ($\pol$) verifies with probability $\Pr[ \texttt{Verify}(\pol(l_x,l_y)) = 1 ] \ge 1 - \mathsf{negl}(\lambda).$ It is \emph{sound} if, for any PPT adversary $\mathcal{A}$ and any $l' \neq l$, $\Pr[ \texttt{Verify}(\pol') = 1 \ \wedge\ \mathsf{Loc}(\pol') = l' ] \le \mathsf{negl}(\lambda)$ unless $\mathcal{A}$ is physically present at $l'$. Proofs are bound to spatio-temporal context, non-transferable, and non-replayable via cryptographic commitments and signatures, resisting relay, distance, mafia, and hijacking attacks.
\end{definition}\vspace{-1mm}

\vspace{-2.5mm}
\begin{definition}[Counter-DoS] \label{def:counterdos} 
A system is DoS-resilient if for every PPT adversary $\mathcal{A}$ issuing at most $q(\lambda)$ queries, the probability that $\mathcal{A}$ causes unavailability, defined as delaying any honest request beyond a fixed bound $\Delta$, without expending computational cost $\Omega(q(\lambda)\cdot \tau)$ is at most $\mathsf{negl}(\lambda)$, where $\tau$ is the per-query puzzle cost. Formally, $\Pr\big[ \mathsf{Delay}(\mathcal{A}) > \Delta \ \wedge\  \mathsf{Cost}(\mathcal{A}) < q(\lambda)\cdot\tau \big] \le \mathsf{negl}(\lambda)$, with $\tau$ enforced via rate-limiting, client puzzles, and authenticated queries.
\end{definition}

\subsection{Scope of Our Solution} 
The $\qpadl$ framework is designed to protect client privacy during the spectrum-query and access phases. In particular, it aims to preserve client anonymity and location privacy while resisting location-spoofing attempts and mitigating DoS attacks in a PQ setting. These protections apply during spectrum querying, cryptographic puzzle retrieval, and subsequent service requests. Our scope is limited to the query and access stages of the SAS workflow. We do not address privacy or security issues arising during user registration, long-term identity management, or actual spectrum utilization after authorization. PU privacy is also outside the scope of this work. In addition, $\qpadl$ does not address location leakage during ongoing spectrum usage, user mobility or handover scenarios~\cite{gao2012location}, nor does it claim protection against side-channel leakage, timing analysis, or physical-layer localization techniques like triangulation~\cite{bahrak2014protecting}. Finally, our proof mechanism provides \emph{proximity-backed location assurance} within the assumed system model and is not intended to constitute a complete defense against all real-world RF-layer localization or relay adversaries.
\section{\textbf{The Proposed Framework: $\qpadl$}} \label{sec:proposedscheme} 
We outline the initial setup of $\qpadl$ framework, followed by its main operations, detailed algorithmic descriptions, and various instantiations and optimization strategies.

%
%
\subsection{$\qpadl$ Framework Initial Setup} \label{subsec:archinitialsetup} 
The initial setup of $\qpadl$ establishes the geolocation DBs, configures APs, and initializes the $\pqtor$ network.

\textbf{Database Structure and Setup:}  
The $\db$ structure in $\qpadl$ follows FCC spectrum-sharing requirements, with each $\psd$ maintaining synchronized entries indexed by a subroutine $\dbindex(\cdot)$ that maps tuples $((l_x,l_y),\ch,\tv)$, consisting of grid-based location coordinates, spectrum channel, and validity window, to specific database blocks. The database is modeled as an $r_\db \times s_\db$ matrix, where each row represents a $b$-bit block partitioned into $s_\db$ words over $GF(2)$ (or, more generally, $GF(2^w)$ depending on the PIR instantiation). In addition to location, indexing may incorporate device characteristics such as transmission power, antenna height, and device category, as well as access preferences including bandwidth and requested duration, thereby supporting accurate and policy-compliant spectrum decisions. This design is consistent with FCC requirements for CBSD geolocation accuracy, which impose bounded horizontal and, for certain device classes, vertical positioning constraints to preserve reliable spectrum assignment and cross-database consistency. The indexing mechanism may be instantiated using FCC-compatible spatial encodings such as geohash, H3, or S2 geometry to support efficient spatial lookup~\cite{agarwal2022survey,jing2025geohash}. After resolving the appropriate index, the $\db.\texttt{Record}$ function stores the associated puzzle and its signature in the corresponding database entry.


\smallskip
\textbf{Access Points Setup:}  
APs in a region collectively form a ring $R_\ap$ containing the public keys of all participants, where each AP holds a key pair $(\sk_\ap, \pk_\ap)$ generated by the FCC using $\lrs.\keygen(pp)$ algorithm (Definition~\ref{def:LRS}). They periodically broadcast time-sensitive beacons ($\beta_\tw$) for device discovery within a time window ($\tw$). The beacon $\beta_{\tw}$ is a region-scoped, time-dependent token without AP-specific identifiers. To derive a confidence-bounded proximity decision, an AP evaluates signal strength and round-trip time using environmental parameters~\cite{robinson2005received, koo2010localizing}, computing the distance as $\Delta \gets \texttt{ProxVerif}(\rss, \rtt, \envv_{\textit{params}})$.



\smallskip
\textbf{$\pqtor$ Setup and Configuration:}   
We assume that PQ-Tor has been initialized and is ready for use via $\pqtor.$ $\setup$. The client proceeds by establishing an anonymous communication circuit using $\pqtor.\circuit$, followed by data transmission through $\pqtor.\texttt{Send}$ and $\pqtor.\texttt{Receive}$, as detailed in Section \ref{subsubsec:pqtor}.

%
%
\subsection{$\qpadl$ Framework Main Operations} 

The overall flow of $\qpadl$ is presented in Algorithms~\ref{Alg:QPADL}-\ref{Alg:PoLAP}, detailing its core operations as follows:

\begin{algorithm}[ht!]
	\small
	\caption{$\qpadl$ Framework}\label{Alg:QPADL}
	\hspace{5pt}       
        \begin{algorithmic}[1]
        \Statex \vspace{-2mm}\hspace{-5mm}\squarenum{1} \underline{$\db \as \psd.\texttt{Puzzle.Bind}(\db)$}:\vspace{+1mm}
        \ForAll{$(l_x, l_y)$}
        \ForAll{$\ch$}
        \ForAll{$\tv$}
    	\State Given $\theta \leftarrow \dbindex((l_x, l_y), \ch, \tv)$
        \State $\pi_{\theta} \as \puzgen(1^\lambda, \kappa)$ 
        \State $\sigma_{\pi_{\theta}} \as \mldsa.\sign(\sk_\psd, \pi_{\theta})$
        \State $\db.\texttt{Record}(\pi_{\theta}, \sigma_{\pi_{\theta}})$
        \EndFor
        \EndFor
        \EndFor \vspace{-1mm}
        \algrule
        \Statex \hspace{-5mm}\squarenum{2} \underline{$\{q_i\}_{i=1}^{n} \as \client.\texttt{Spectrum}.\query((l_x,l_y), \ch, \tv, \tw, \beta_\tw)$}:\vspace{+1mm}
        \State $(C_\tw, \sigma_\ap, e_\id) \as \client.\pol((l_x, l_y), \tw, \beta_\tw)$
        \State $\theta \as \dbindex((l_x, l_y), \ch, \tv)$
        \State $(q_1, q_2, ..., q_n) \as \pir.\query(\theta)$
        \For{$i = 1, \dots, n$}
        \State $\pqtor.\texttt{Send}(\psd_i; q_i, C_\ts, \sigma_\ap, e_\id)$
        \EndFor \vspace{-1mm}
        \algrule \vspace{-1mm}
        \State $\psd.\pqtor.\texttt{Receive}(\su; q_i, C_\tw, \sigma_\ap, e_\id)$
        \Statex \hspace{-5mm}\squarenum{3} \underline{$\rho_i \as \psd.\texttt{Spectrum}.\response(q_i, C_\tw, \sigma_\ap, e_\id, R_\ap)$}:\vspace{+1mm}
        \If{$1 = \pol.\verify(e_\id, \sigma_\ap, C_\tw, R_\ap)$}
        \If{$1 = \lrs.\link(e_\id, \sigma_\ap, C_\tw, R_\ap, \vec{\sigma}_\ap)$},~\Return $\perp$
        \Else
        \State Record $\sigma_\ap$
        \State $\rho_i \as \pir.\query.\response(q_i, \db)$
        \EndIf
        \EndIf
        \State $\psd.\pqtor.\texttt{Send}(\su; \rho_i)$ \vspace{-1mm}
	\algrule\vspace{-1mm}
        \For{$i = 1, \dots, k$, where $k\in[t, n]$}
        \State $\client.\pqtor.\texttt{Receive}(\psd_i; \rho_i)$
        \EndFor
        \State $\db_{\theta} \as \client.\pir.\blockrec(\theta, (\rho_1, \rho_2, ..., \rho_k))$
        \Statex \hspace{-5mm}\squarenum{4} \underline{$Token \as \client.\createtoken(\Pi, \sigma_{\Pi})$}:
        \If{$1 \as \mldsa.\verify(\pk_\psd ,\Pi, \sigma_\Pi)$}
        \State $\Psi \as \pow(\Pi, \kappa)$
        \State $\Return~\textit{Token} \leftarrow (\Pi, \sigma_\Pi, \Psi)$  
        \EndIf \vspace{-1mm}
        \algrule\vspace{-1mm}
        \Statex \hspace{-5mm}\squarenum{5} \underline{$\{0,1\} \as \client.\service.\request(C_\tw, \sigma_\ap, e_\id)$}:\vspace{+1mm}
        \If{$1 = \mldsa.\verify(\pk_\psd, \sigma_{\Pi})$,}
        \If{$1 = \powverif(\Pi, \Psi)$},~Record the $\textit{Token}$
        \If{$1 \as \pol.\verify(C_\tw, e_\id, \sigma_\ap, R_\ap)$,} 
        \State $\Return~1$, and grant access.
        \EndIf
        \EndIf
        \EndIf 
        \end{algorithmic}
\end{algorithm}
\setlength{\textfloatsep}{0pt}

\textbf{1) Puzzle Management and Private Spectrum Services:}  
Each $\psd$ initializes and maintains a synchronized spectrum DB containing spectrum-availability information together with pre-generated cryptographic puzzles and their signatures. For each admissible DB index, derived from location coordinates, channel, validity window, and relevant device attributes (e.g., power level, height, or category), the $\psd$ generates a puzzle at the prescribed difficulty level and signs it using $\mldsa.\sign$. The resulting records are stored under $\dbindex((l_x,l_y), \ch, \tv)$ and refreshed according to their validity interval (e.g., hourly). The number of puzzles generated depends on device characteristics, server count, and their maximum capacity~\cite{agarwal2022survey}. This setup enables clients to retrieve both spectrum information and a signed anti-DoS puzzle through the same privacy-preserving query process.


\textbf{2) Proximity-Backed Location Assurance and $\pol$ Generation:}  
Before issuing a spectrum query, the client obtains a fresh proof artifact for the current time window $\tw$ from a nearby AP, as shown in Algorithm~\ref{Alg:PoLAP}. Upon receiving the latest beacon $\beta_\tw$, the client computes a commitment $C_\tw$ to its location, the current window, and a random nonce, and sends it to the selected AP. The AP first checks beacon freshness and then applies a proximity-verification procedure based on local signal measurements (e.g., RSS and RTT) to determine whether the requester is plausibly within the AP’s vicinity. If this check succeeds, the AP derives an event identifier $e_\id$ from the AP ring, the current beacon, and the active time window, and produces a ring signature over the client’s commitment. The client verifies this signature and accepts $(\sigma_\ap,e_\id)$ as a valid proof artifact for subsequent query admission.



\begin{algorithm}[ht!]
	\small
	\caption{${(C_\ts, \sigma_\ap, e_\id) \as \pol((l_x,l_y), \ts, \beta_\ts)}$}\label{Alg:PoLAP}
	\hspace{5pt}\vspace{-1mm}
                \begin{algorithmic}[1]
        \Statex \hspace{-5mm}\squarenum{1} \underline{$C_\ts \as \client.\pol.\request(\beta_\tw)$}:\vspace{+1mm}
        \State Given the latest beacon $\beta_\tw$:
        \State $r_c \asrandom \zq$
        \State $C_\tw \as H((l_x, l_y) || \beta_\ts || \tw || r_c)$
        \State Send $(\beta_\tw, C_\ts)$ to $\ap$. \vspace{-1mm}
        \algrule \vspace{-1mm}
         \Statex \hspace{-5mm}\squarenum{2} \underline{$(\sigma_\ap, e_\id) \as \ap.\pol.\response(\beta_\tw, C_\tw)$}:\vspace{+1mm}
        \If{$\beta_\ts$ is not the latest beacon:},~\Return $\perp$
        \Else 
        \State $\Delta_c \as \ProxVerif(\rss, \rtt, \envv_{\textit{params}})$
            \If{$\Delta_c \leq \Delta_{th}$}
                \State $e_\id \as H'(R_\ap, \beta_\tw, \tw)$
                \State $\sigma_\ap \as \lrs.\sign(e_\id, \sk_\ap, C_\ts, R_\ap)$
            \EndIf
        \EndIf
        \State Send $(\sigma_\ap, e_\id)$ to the Client. \vspace{-1mm}
	\algrule\vspace{-1mm}
        \Statex \hspace{-5mm}\squarenum{3} \underline{$\{(\sigma_\ap, e_\ap), \perp\} \as \client.\pol.\verify(\sigma_\ap, e_\id, C_\ts, R)$}:\vspace{+1mm}
        \If{$1 = \lrs.\verify(e_\id, \sigma_\ap, C_\ts, R)$},
        \State \Return $(\sigma_\ap, e_\ap)$
        \EndIf 
    \end{algorithmic}     
\end{algorithm} 
\setlength{\textfloatsep}{0pt}

\textbf{3) Private Spectrum Query and Puzzle Retrieval:}  
After obtaining a valid proof artifact (Step 8), the client computes the target database index $\theta \gets \dbindex((l_x,l_y),\ch,$ $\tv)$ and generates PIR queries for the participating $\psd$s (Steps 9-10). These queries are transmitted through $\pqtor$ together with the committed proof tuple $(C_\tw,\sigma_\ap,e_\id)$, thereby combining query privacy with communication unlinkability (Steps 11-12). Upon receiving a request, each $\psd$ verifies the proof artifact and checks whether the submitted proof is linkable to a previously accepted proof under the same event identifier and time window. If the proof is linked, the request is rejected according to the one-use-per-window policy; otherwise, the proof is recorded and the $\psd$ returns its PIR response (Steps 15-19). Once sufficiently many responses are collected, the client reconstructs the requested database block and extracts the corresponding spectrum information, puzzle, and puzzle signature. In this way, the same event-scoped proof supports both privacy-preserving query admission and lightweight rate limiting against repeated or abusive requests.


\textbf{(4) Token Creation and SAS Service Request:}  
After reconstructing the target DB entry, the client verifies the authenticity of the retrieved puzzle using the $\psd$'s public key and then solves the puzzle to generate a service token (Steps 23-24). The token is subsequently submitted when requesting SAS-related services. Upon receiving the request, the server verifies the puzzle signature, checks the validity of the puzzle solution, and records the token to prevent replay or repeated computational abuse (Step 26). It then verifies the accompanying proof artifact to confirm that the request is bound to a fresh, proximity-backed commitment generated within the valid time window (Steps 27-28). If all checks succeed, the server requests the client to reveal the commitment and checks $C_\tw = H((l_x, l_y) || \beta_\tw || \tw)$ and grants service access. This phase ensures that service requests are tied to both a valid spectrum-query outcome and a computationally bounded anti-DoS mechanism.


%
%
\section{$\qpadl$ Framework Instantiations}\label{sec:instan1} 
This section presents concrete realizations of $\qpadl$ under different deployment and performance objectives. We first instantiate the query and service phases using specific PIR- and PoW-based constructions, then describe parallelization strategies for improving online response latency, and finally discuss practical optimizations that further improve deployability and efficiency.

%
%
\subsection{Cryptographic Primitive Instantiation} \label{subsec:primitiveinstantiation}
We instantiate $\qpadl$ using concrete building blocks for the spectrum-query and service-request phases. In particular, we present three PIR-based query instantiations and two PoW-based service instantiations, each capturing a different trade-off among privacy assumptions, robustness, communication, and online efficiency, followed by a complete end-to-end realization shown in Figure~\ref{fig:instantiation}.

\subsubsection{PIR Instantiations} 
While any $t$-private PQ-secure PIR can be employed in $\qpadl$, we introduce three tailored instantiations, each offering distinct security features and trade-offs. 

\noindent{\em \underline{(i) $\qpadlenc$}:} This is the most efficient instantiation for the query phase, based on the IT-secure PIR scheme from \cite{chor1998private}, which assumes $n$ non-colluding, responsive servers maintaining synchronized database copies, hence named ENS (Efficient Non-colluding Servers). The client performs only XOR operations over random $r$-bit vectors, while each $\psd$ performs a single multiplication of the query over the entire database matrix. The PIR algorithm is described in \cite{chor1998private}, with a GPU-parallelized version detailed in Section \ref{subsubsec:chorpir}. This setup is best suited to settings where strict non-colluding synchronized $\psd$s is a reasonable deployment assumption and minimum client-side overhead is desired.  

\noindent{\em \underline{(ii) $\qpadlftr$}:} This Fault Tolerant Robust (FTR) instantiation of $\qpadl$ adopts the IT-secure PIR scheme from \cite{goldberg2007improving}, enabling $\nu$-Byzantine fault tolerance by allowing block reconstruction even if up to $\nu$ servers return incorrect responses. Servers compute a query vector multiplication over the DB matrix. Using Shamir’s secret sharing, the client can reconstruct the target block via Lagrange interpolation when responses are received from any $k$ out of $[t, n]$ $\psd$s. In cases of synchronization errors, transmission faults, or Byzantine behavior (${\nu < k}$), the client employs Guruswami-Sudan list decoding for error correction, ensuring robustness. Further details of this PIR are provided in \cite{goldberg2007improving}, with its parallelized implementation described in Section \ref{subsubsec:goldbergpir}. Compared with $\qpadlenc$, this instantiation trades additional reconstruction complexity for stronger resilience to faulty, inconsistent, or Byzantine $\psd$ responses. 


\noindent{\em \underline{(iii) $\qpadloop$}:} This instantiation, named for its use of Online-Offline Preprocessing (OOP), maximizes efficiency and enhances DB structure while providing computational security. It employs CIP-PIR~\cite{gunther2022gpu}, a PQ-secure PIR protocol optimized for large-scale networks with multi-GB DBs where servers hold identical but not strictly replicated data. Improving upon Chor et al.~\cite{chor1998private}, it uses seed-based queries to reduce communication and restricts each server’s data access to a subset, minimizing online computation. Unlike Chor’s model, seed selection occurs server-side during preprocessing. CIP-PIR includes two preprocessing steps: a one-time DB preprocessing during setup and a client-independent step that precomputes response components unrelated to any specific query, enabling reuse for a single future query. This setup assumes $\psd$s maintain the same data, though chunk order may vary.  Due to its favorable online efficiency and communication profile, $\qpadloop$ serves as the representative query instantiation in our end-to-end workflow of Figure~\ref{fig:instantiation}. Its use reflects a deployment-oriented design choice for large-scale SAS settings, while explicitly shifting from information-theoretic privacy guarantees to computational security.




\subsubsection{\textbf{PoW Instantiations}} The service-request phase of $\qpadl$ can be instantiated with different PQ-secure PoW mechanisms depending on the desired trade-off between client-side cost, verification efficiency, and communication overhead.

\noindent {\em \underline{(i) $\qpadlhct$}:} Given the burden of token creation on the users, this instantiation adopts the Hashcash Tree ($\hct$) construction to provide an efficient and PQ-secure PoW mechanism, considering users' diverse computational resources (e.g., CPU, energy, battery, storage). As an enhanced version of the original Hashcash puzzle used in Bitcoin~\cite{aura2000resistant}, $\hct$ is designed to resist quantum and parallel brute-force attacks~\cite{chiriaco2017finding, alviano2023hashcash}, offering adjustable difficulty tailored to varying user capabilities. Built on hash functions, it inherently ensures efficiency in both classical and PQ settings, leveraging sequential hardness via a binary tree structure that prevents full parallel shortcuts. For this reason, $\qpadlhct$ is the practical default in our full instantiation, especially when user devices have heterogeneous computational and energy budgets. The detailed $\hct$ algorithm is presented below: \vspace{-4mm}

\begin{itemize}[leftmargin=*]
    \item[-] $\Pi \as \hct.\puzgen(1^\lambda, \kappa)$: Randomly selects $n_s \asrandom \{0,1\}^\lambda$ and sets the number of leaves $n_l$ based on the difficulty level $\kappa$. The puzzle is $\Pi = (h, n_s, \kappa, n_l)$, where $h$ is the hash function and $n_s$ is the selected randomness. \vspace{-2mm}
    
    \item[-] $\Psi \as \hct.\pow(\Pi)$: Constructs a perfect binary tree of Hashcash puzzles via brute force. For each leaf node ($i > n_l$), the client finds $n_x$ such that $h_\kappa(n_s || i || 0 || 0 || n_x)$ has $\kappa$ leading zeros. For internal nodes ($i < n$), it solves $h_\kappa(n_s || i || h_{2i} || h_{2i+1} || n_x)$. The root nonce $n_1$ serves as the PoW token, with nonces $(n_{2^{n_l}-1}, \dots, n_1)$. \vspace{-2mm}

    \item[-] $\{0,1\} \as \hct.\powverif(\Pi, \Psi)$: Upon receiving the root's nonce, the server randomly selects a leaf index $i \in [1, n_l]$ and requests the corresponding path from leaf to root. Verification requires only $\log n$ hash computations. \vspace{-1mm}
\end{itemize}

\noindent {\em \underline{(ii) $\qpadllbp$}:} An alternative PoW instantiation for $\qpadl$ leverages the lattice-based construction from \cite{behnia2021lattice}, built on the hardness of the Shortest Vector Problem (SVP) \cite{SVPChallenge}. This alternative offers security rooted in lattice hardness, but at a noticeably higher computational and communication cost than hash-based constructions, making it more suitable for settings where such overhead is acceptable. It comprises three core subroutines, described below: 
\vspace{-2mm}
 
\begin{itemize}[leftmargin=*] 
    \item[-] $\Pi \leftarrow \lb.\puzgen(1^\lambda, 1^{n_\Lambda}, \kappa)$: Given security parameter $\lambda$, lattice dimension $n_\Lambda$, and difficulty level $\kappa$, the algorithm samples from uniform distribution $x_2, \dots, x_n \leftarrow \mathcal{U}({0} \cup [p-1])$, sets $\alpha = 1.05 \cdot \Gamma(n/2+1)^{1/n}/\sqrt{\pi}$, and constructs the lattice basis which is an $n_\Lambda \times n_\Lambda$ matrix $B$ with the first row comprised of $[p~x_2~\dots~x_n]$, ones on the subdiagonal, and zeros elsewhere. It outputs the puzzle $\Pi = (\alpha, n_\Lambda, B, p)$. \vspace{-2mm}
    
    \item[-] $\Psi \as \lb.\pow(\Pi, \kappa)$: Solves the puzzle $\Pi$ by finding $v \in \Lambda(B)$ such that $||v|| \leq \alpha \cdot p^{1/{n_\Lambda}}$, and returns $\Psi = (v, \nu)$ where $v = B \cdot \nu$. \vspace{-2mm}

    \item[-] $\{0,1\} \as \lb.\powverif(\Pi, \Psi)$: Verifies the solution by checking $||v|| \leq \alpha \cdot p^{1/{n_\Lambda}}$, $v = B \cdot \nu$, and $\nu \in \mathbb{Z}^{n_\Lambda}$, returning $1$ if all conditions hold. 
\end{itemize}

\paragraph*{Instantiation Rationale and Selection.}
The $\qpadl$ framework is designed to support multiple instantiation choices because different SAS deployments prioritize different goals, such as stronger information-theoretic query privacy, robustness to faulty or unresponsive DBs, or reduced online computation at scale. Accordingly, the PIR layer governs how spectrum information and signed puzzles are retrieved under different trust and performance assumptions, while the PoW layer governs how service requests are rate-limited under different client-side resource constraints. These instantiations are not independent add-ons; rather, they represent interchangeable design points within the same $\qpadl$ workflow, enabling the framework to be adapted to federated, robust, or large-scale deployment settings. Further details about the performance of each instantiation are provided in Section~\ref{sec:PerfEval}.

\subsubsection{\textbf{Full Instantiation of $\qpadl$}} 
\begin{figure}
	\includegraphics[width=\linewidth] {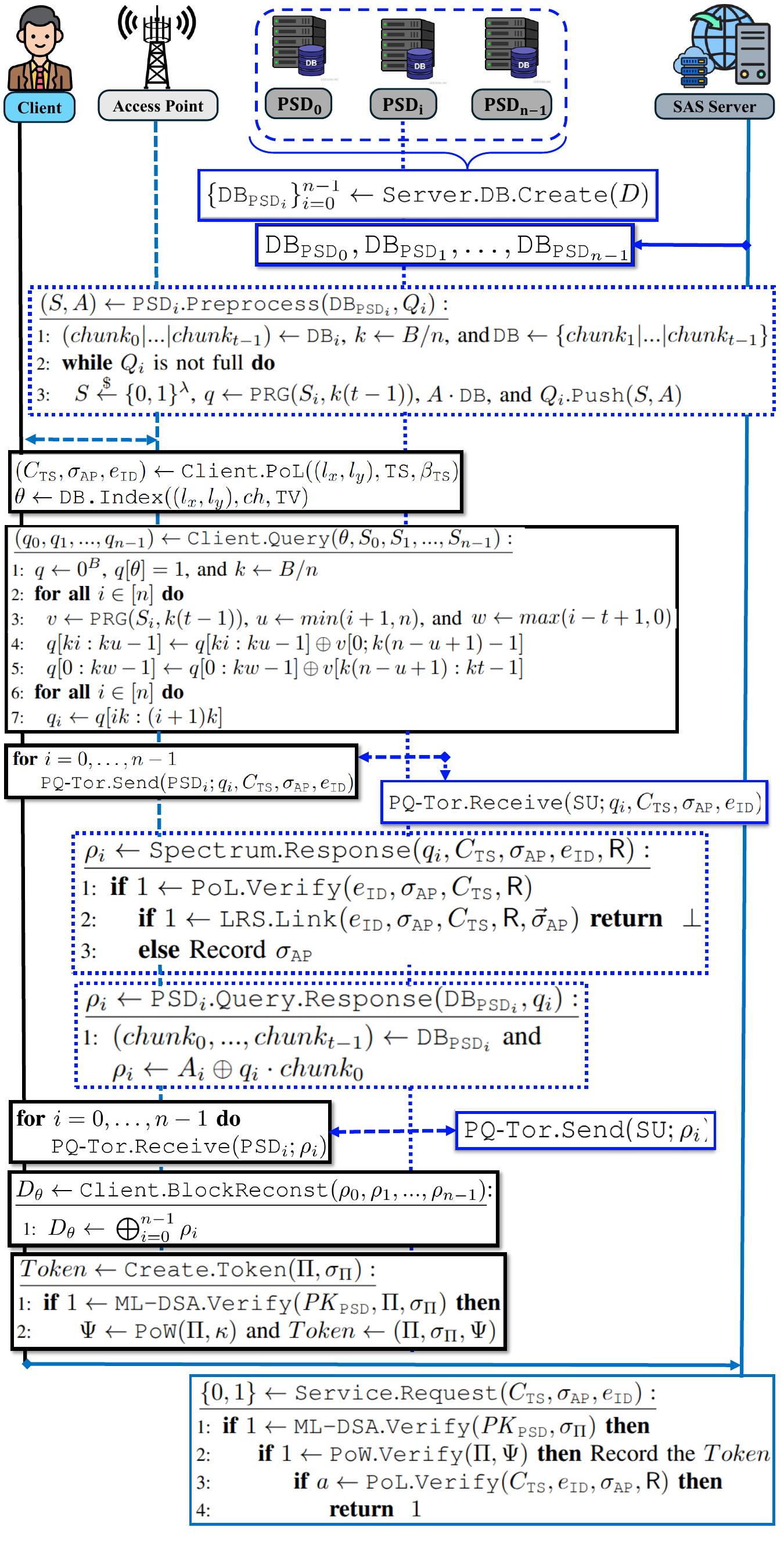}  
	\caption{$\qpadl$ Full Instantiation} 
	\label{fig:instantiation}
\end{figure}

Figure~\ref{fig:instantiation} illustrates a complete realization of $\qpadl$ using $\qpadloop$ for privacy-preserving spectrum queries and $\qpadlhct$ for service-side request throttling. This instantiation is chosen as a representative deployment-oriented design point because it combines reduced online $\psd$ computation with efficient client-verifiable PoW, while preserving the overall $\qpadl$ workflow and security goals. The $\db$ contains frequency data $((l_x, l_y), ch, \tv)$, with each entry bound to a signed puzzle $(\pi_\theta, \sigma_{\pi_\theta})$ via $\texttt{Puzzle.Bind}()$ (Algorithm\ref{Alg:QPADL}, Steps 1-7). While all $\psd$s store the same content, their chunk orders may differ. Each block $D_\theta = ((l_x, l_y), ch, \tv, \Pi_\theta, \sigma_{\Pi_\theta})$ is divided into $B$ chunks, with $k = B/n$. For all $i \in [n]$, chunks are defined as $chunk_i = block_{ki}|...|block_{ki+k-1}$ and each $\psd$ arranges its DB as $\db_{\psd_i} = chunk_i|...|chunk_{i+t-1\bmod n}$.

The CIP-PIR scheme requires each $\psd_i$ to perform a one-time, client-independent preprocessing on its local DB, generating seed-value pairs $(S, A)$ and storing them in a local queue $Q_i$. Specifically, for each DB chunk, a randomly chosen seed $S$ is expanded into a $k(t-1)$-bit query $q$ using a pseudo-random generator ($\texttt{PRG}$) (Steps 1-2). The corresponding value $A$ is computed by XORing all non-flip chunks whose positions in $q$ are set to $1$, and the resulting $(S, A)$ pair is added to $Q_i$ (Step 3). Since each $\psd$ holds $(t-1)$ non-flip chunks out of $n$, the preprocessing phase covers $(t-1)/n$ of the DB, leaving only $1/n$ for computation during the online phase. 

To access spectrum, clients must first obtain a valid event-scoped proof artifact in phase two. After receiving the latest beacon $\beta_\tw$ from a nearby AP,  
the client sends a location commitment as a $\pol$ request. If the AP’s proximity check succeeds, it derives an event identifier from the current beacon, AP ring, and active time window, signs the client’s commitment, and returns $(\sigma_\ap, e_\id)$ (Algorithm~\ref{Alg:PoLAP}, Steps 1-12). The client then computes the index via $\dbindex$ and proceeds with the PIR query. It constructs a zero-initialized $b$-bit vector with the $\theta$-th bit set, expands each $\psd$’s seed $S_i$ into a $k(t-1)$-bit vector $v$ using a $\texttt{PRG}$, and XORs $v$ into the corresponding chunks of the query. The final query is partitioned into $n$ sub-queries $q_i$, each mapped to the designated chunk of server $\psd_i$ (Steps 1-7). Finally, the client privately transmits the proof artifact and queries $(q_i, C_\tw, \sigma_\ap, e_\id)$ to each $\psd_i$ through $\pqtor$.

Upon receiving a query, each $\psd$ validates $\pol$ with $\pol.\verify$ (Algorithm~\ref{Alg:QPADL}, Steps 14-15) and applies $\lrs.\link$ to detect reuse within the same time window. Replayed proofs are rejected; otherwise, the proof is logged and the PIR response is generated. Thus, the same event-scoped proof artifact serves two purposes at query time: it admits only fresh, proximity-backed requests and enforces one-use-per-window rate limiting through linkability under the same event identifier. Next, $\psd_i$ retrieves its assigned flip chunk (e.g., $chunk_0$ in $\db_i$), XORs the blocks indicated by set bits in $q_i$, and combines the result with the precomputed value $A_i$ from the offline phase. Since only one chunk ($1/n$ of the DB) is accessed online, with the remaining $(t-1)/n$ processed beforehand, computation is minimized. The response is then sent back via the established $\pqtor$.

Upon receiving responses, the client reconstructs the target block via $\blockrec$ and verifies the $\mldsa$ signature on the retrieved puzzle. It then solves the puzzle with $\hct.\pow$, derives the root solution $\psi$, and forms the token $Token \as (\Pi, \sigma_\Pi, \Psi)$. To access a SAS service, the client submits its location commitment, $\pol$, and the token. The server verifies the $\psd$’s puzzle signature and the PoW solution; invalid or reused solutions are rejected as DoS attempts. Valid tokens are logged to prevent replay, and access is granted after confirming the proof of location.  
In certain services or when client behavior is suspicious, the server may request disclosure of the committed location, assumed to occur over a secure authenticated channel (e.g., PQ-TLS), which is beyond this work’s scope.

%
%
\subsection{Instantiation via Parallelization} \label{subsec:parallelization} 
Our parallelization objective is to reduce online $\psd$ response latency under concurrent user demand, since this component directly affects the practicality of privacy-preserving spectrum access in large-scale SAS deployments. Numerous GPU-accelerated works \cite{gunther2022gpu, lam2023gpu, li2025cat} have been proposed to improve the efficiency of PIR schemes. Gunther et al. \cite{gunther2022gpu} propose CIP-PIR, where GPU acceleration is leveraged to improve offline server computation, but without addressing online server runtime, which is directly related to response delay. \cite{li2025cat} attempts to improve the efficiency of homomorphic encryption-based single-server PIR and therefore does not align with our design rationale to provide resiliency against rogue attacks and a single root of trust. 
We next summarize the relevant GPU execution model and then present parallelized realizations of the PIR response computation

\textbf{NVIDIA GPU Architecture and CUDA:} \label{subsubsec:CUDA}
A Graphical Processing Unit (GPU) is designed to accelerate computationally intensive tasks by leveraging Single Instruction Multiple Threads (SIMT) execution. An NVIDIA GPU comprises multiple Stream Multiprocessors (SMs), each of which manages several CUDA cores. The latter executes general-purpose computations, where each operates at a base clock frequency (e.g., 1320 MHz). The circuit provides hierarchical memory types: 
{\em (i) global memory ($\approx$GBs)}: can be accessed by all threads in SMs. It is an off-chip memory and the data transfer medium between system memory and GPU with high access throughput. 
{\em (ii) shared memory ($\approx$KBs)}: is shared among cores in a single SM and provides a faster memory access compared to (i). 
{\em (iii) registers}: reside in each core, with the fastest memory access to hold the frequently accessed and local data.
CUDA is an interface developed by NVIDIA \cite{kirk2007nvidia}, which allows programmers to define and execute operations on GPUs. 
A CUDA program launches a {\em kernel} over a multidimensional {\em grid} of blocks. Each {\em block} contains multiple {\em warps} (that is, the warp comprises 32 threads) running under the SIMT paradigm.

\textbf{Parallel Chor-PIR:} \label{subsubsec:chorpir} The online $\psd$ computational overhead consists of matrix multiplication over $GF(2)$ (i.e., $\rho \as q \cdot \db$), where $\rho \in \{0,1\}^{b},~\db \in \{0,1\}^{r \times b}$, and $q \in \{0,1\}^{r}$. 
This operation consists of conditional aggregation (via bitwise XOR) of matrix rows $\db$ based on the query's bit values $\{q_{r'}\}_{r' \in \{1,\ldots r\}}$. 
The total computational work per query is quantified as $\mathbb{W} = nnz(q) \cdot b $, where $nnz(q)$ denotes the Hamming weight of the query. 
The total memory traffic is equal to $\mathbb{Q} \as r + nnz(\rho) \cdot b + b$, accounting for the load of $\rho$ and active rows of $D$, as well as the store of the response $R$. 
Consequently, the server-side computation is a memory-bound task due to its low operational intensity, formalized as $\mathbb{I}=\frac{\mathbb{W}}{\mathbb{Q}}$. Zhang et al. \cite{zhang2025can} demonstrate that Boolean linear algebra is not suited for GPU tensor cores. Instead, optimization efforts must be pivoted towards CUDA cores using memory hierarchy exploitation, data locality, and coalesced memory access.

Our proposed parallel Chor-PIR is detailed in Algorithm \ref{Alg:GPU-chor}. $\query.\response$ processes mutliple queries $\qvec = \{q_i\}_{i=1}^{\qlen}$ in batches. First, it copies the queries $\qvec$ to global memory before invoking the CUDA kernel (Step 2). The $\db$ is already residing in the global memory. The kernel is configured with $(\qlen, \lfloor \frac{b}{32} \rfloor)$ blocks, each is configured as a $(32,8)$ thread layout to enable warp-level parallelism and strided row-wise accumulation. 
Each thread $(t_x,t_y)$ computes partial $GF(2)$ XOR accumulations on rows using an 8-stride loop. In particular, Step 9 introduces warp divergence due to the conditional row fetch based on $q_{b_x,r'}$.
However, empirical profiling shows that the high memory throughput of this memory-bound kernel compensates for this divergence. One can use a bitmasking approach to avoid this divergence, but it incurs inferior performance, especially when $q_i$ is a sparse vector (i.e., $nnz(q_i) \ll r$). 
The intermediate accumulations are written to shared memory for intra-level reduction. Threads with $ty=0$ finalize the response by aggregating the 8 per-thread vertical sub-sums $\{a_{t_x,r'}\}_{r'}$, and committing the result to the global memory.

\begin{algorithm}[h]
	\small
    \caption{Multi-request Parallel Chor-PIR}\label{Alg:GPU-chor}
    \begin{algorithmic}[1]
        %
        \Statex $\underline{\rvec \as \psd.\query.\response(\qvec)}$:
        Upon receiving multiple requests $ \qvec = \{q_i = \{ q_{i,1},q_{i,2},\ldots ,q_{i,r} \}\}_{i=1}^{{\qlen}} \in_R GF(2)^r$
        \State \textbf{each} $\psd_i$ for $i = 1, 2, ...,\ell$ \textbf{do} 

            \State \hspace{1em} copy $\qvec$ from main memory to global memory
            \State \hspace{1em} $|grid| = (|\qlen, \lfloor \frac{b}{32} \rfloor)$ , $|block|=(32,8)$
            \State \hspace{1em} \textbf{CUDA Kernel:}
            \State \hspace{2em} \textbf{for each} block $(b_x,b_y) \in \{(1,1) , \ldots, |grid|\}$ \textbf{do} 
            \State \hspace{3em} \textbf{for each} thread $(t_x,t_y) \in \{(1,1) , \ldots , |block| \}$ \textbf{do} 
            \State \hspace{4em}  $c \as 32 \cdot b_y + t_x$ , $a \as 0$
            \State \hspace{4em}  \textbf{for} $r' = t_y, \ldots, r~[\textbf{step}=8]$ \textbf{do}
            
            \State \hspace{5em} \textbf{if} $\qvec_{b_x,r'} = 1$ \textbf{then} $a \as a + \db_{r',c}$
            \State \hspace{4em} store $a$ into shared memory: ${a}_{t_x,t_y} \as a$
            \State \hspace{4em} synchronize threads in the block $(b_x,b_y)$
            \State \hspace{4em} \textbf{if} $t_y = 0$ \textbf{then} 
            \State \hspace{5em} $\rho_{b_x,c} \as \bigoplus_{r'=1}^8 a_{t_x,r'}$
            \State \hspace{5em} write $\rho_{b_x,c}$ from register to global memory
        \State \hspace{1em} \Return $\rvec = \{ \rho_i = \{ \rho_{i,1} , \ldots, \rho_{i,b} \} \}_{i=1}^{\qlen}$
    \end{algorithmic}
\end{algorithm}

\textbf{Parallel Goldberg-PIR:} \label{subsubsec:goldbergpir}
Algorithm~\ref{Alg:GPU-goldberg} outlines the GPU-accelerated implementation of the Goldberg PIR. It operates in a batched setting, each containing a vector of queries $\qvec$, each processed over a $\db$. $\qvec$ and $\db$ are of size $\qlen \cdot r$ and $r \cdot b$, respectively. For each incoming PIR query batch $\qvec$, the host transfers the corresponding matrix to the global memory. Tiling parameters are then set to configure the kernel launch dimensions. The algorithm adapts the tile sizes ($bm$, $bn$) based on the query and DB dimensions to exploit thread utilization and memory locality (Steps 3-4). The CUDA $grid$ and $block$ sizes are determined based on tile sizes to parallelize the workload over matrix dimensions (Steps 5).

Within the CUDA kernel, each thread block computes a tile of the result matrix. 
Each thread performs a series of multiply-accumulate operations over the shared memory tiles to compute partial results for its assigned output element. These partial results are stored in thread-local registers and accumulated across all tile iterations (Steps 11-16). After full computation loop over the depth dimension $r$ completes, each thread writes its final result to the output buffer $\rho$ in global memory (step 17). Upon completion of the kernel execution, the host retrieves the response matrix $\rvec$ from the device memory and finally returns it as an output.  
The algorithm leverages several GPU optimization strategies inspired by best practices in high-performance matrix multiplication \cite{kelefouras2016high} such as shared memory caching, register blocking, warp-level parallelism, and conditional adjustment of tile sizes. These techniques collectively enhance arithmetic intensity and scalability, enabling practical deployment of PIR at scale.

\begin{algorithm}[h]
	\small

    \caption{Multi-request Parallel Goldberg-PIR}\label{Alg:GPU-goldberg}
    \begin{algorithmic}[1]
        \Statex $\underline{\boldsymbol{\rho} \gets \psd.\query.\response(\qvec)}$: Upon receiving multiple PIR requests $\qvec = \{q_i = (q_{i1}, q_{i2}, \ldots, q_{ir})\in \mathbb{F}^r\}^{\qlen}_{i=1} $

        \State \hspace{1em} copy $\qvec$ from main memory to global memory
        \State \hspace{1em} $br = 8$ , $tn = 8$ , $tm = 8$ \Comment{tiling setting}
        \State \hspace{1em} \textbf{if} $\qlen \ge 128$ and $n \ge 128$ \textbf{then} $bm = 128$ , $bn = 128$ 
        \State \hspace{1em} \textbf{else} $bm=64$ , $bn=64$
        \State \hspace{1em} $|grid| = (\lfloor \frac{n}{bn} \rfloor, \lfloor \frac{m}{bm} \rfloor)$, $|block|=(bn, bm)$
        \State \hspace{1em} \textbf{CUDA Kernel:}
        \State \hspace{2em} \textbf{for each} block $(b_x,b_y) \in |grid|$ \textbf{do}
        \State \hspace{3em} \textbf{for each} thread $(t_x,t_y) \in |block|$ \textbf{do}
        \State \hspace{4em} $r_x = tm \cdot b_x + t_x$ , $r_y = tn \cdot b_y + t_y$
        \State \hspace{4em} $\rho_{c_x,c_y} = 0, \forall c_x \in \{1,\ldots tm\} , c_y \in \{1,\ldots,tn\} $
        \State \hspace{4em} \textbf{for} $k = 0 ,\ldots, r~[\text{step}~br]$ \textbf{do}
        \State \hspace{5em} copy $\qvec_{r_x,[k,\ldots,k+br]}$ to shared memory
        \State \hspace{5em} copy $\db_{[k,\ldots,k+br],r_y}$ to shared memory
        \State \hspace{5em} synchronize threads in the block $(b_x,b_y)$
        \State \hspace{5em} $s \as \sum_{i=1}^{br} {\qvec_{r_x,k+i} \cdot \db_{k+i,r_y} \bmod q}$
        \State \hspace{5em} $\rho_{t_y,t_x} \as \rho_{ty,t_x} + s \bmod q$
        \State \hspace{5em} synchronize threads in the block $(b_x,b_y)$
        \State \hspace{4em} copy $\rho_{t_y,t_x}$ to global memory

        \State \hspace{1em} \Return $\rvec = \{ \rho_i = \{ \rho_{i,1} , \ldots, \rho_{i,b} \} \}_{i=1}^{\qlen}$
    \end{algorithmic}
\end{algorithm}

%
%
\subsection{Instantiation Optimizations} \vspace{+1mm}\label{subsec:Optimization}

\textbf{Offline-Online mode:} The offline-online execution mode improves practicality across several phases of $\qpadl$ by shifting reusable computations away from the critical online path. {\em (i)} In the $\pol$ phase, static ring configuration enables reusable preprocessing and reduces repeated online work. {\em (ii)} In the spectrum query phase, the main bottleneck which is the PIR responses that grow linearly with DB size, can be alleviated by offline-online precomputations on $\psd$ servers. {\em (iii)} In the service request phase, the $\hct$-based PoW can be converted to a non-interactive form via the Fiat-Shamir heuristic~\cite{canetti2019fiat}, enabling users to generate verifiable tokens independently without interactive communication.

\textbf{Database Compression Techniques:} Since DB size directly affects $\psd$ computation, particularly in PIR, compression can substantially improve spectrum query performance in $\qpadl$ (see Section~\ref{sec:PerfEval}). $\qpadl$ employs a technique that sorts DB entries (e.g., FCC frequency data) and stores differences between successive items instead of full values~\cite{tamrakar2017circle}. As adjacent entries are often similar, their differences require fewer bits. This method applies to all block-based PIR schemes used in $\qpadl$, reducing storage from $r_\db \times b$ bits to about $O(r_\db(b - \log(r_\db)))$ bits. The resulting efficiency gains in computation and storage are detailed in Section~\ref{sec:PerfEval}.  This optimization reduces storage and response-side processing cost without changing the privacy guarantees of the underlying PIR construction.
 
\textbf{Multiple Block Retrieval:} To reduce repeated query overhead, a single DB block may store multiple puzzles of increasing difficulty, enabling the client to retrieve several service-use tokens within one PIR query. This optimization preserves index indistinguishability because the queried block index remains hidden by the underlying PIR, even when multiple puzzles are embedded in the same block. However, as discussed earlier, retrieving several puzzles at once may reduce future query frequency and thus expose limited coarse-grained temporal behavior at the service layer; it does not reveal which DB block was queried. Accordingly, this optimization improves communication efficiency while preserving query privacy at the PIR layer.


\vspace{+2mm}
\noindent\textit{Overall, these instantiations show that $\qpadl$ is not tied to a single cryptographic realization; rather, it defines a common SAS access workflow adaptable to different privacy, robustness, and efficiency requirements. The chosen end-to-end instantiation represents one operating point, while the broader design space highlights how the framework can be tuned for different deployment environments.}

%
%
\section{Security Analysis} \label{sec:Security} 
We present a sequence of security proofs aligned with the threat, security, and system models. Below, we state the main theorems and lemmas for $\qpadl$; the full analysis and detailed proofs appear in Appendix~\ref{sec:SecuirtyAppendix}. \vspace{-1mm}

\begin{theorem}\label{the:mainQPADL}
For any QPT adversary interacting with $\qpadl$, the framework achieves, except with probability $negl(\lambda)$: \emph{(i) PQ-secure location private spectrum access}, where the privacy of the queried spectrum index reduces to the security of the instantiated PIR layer, namely IT-secure for $\qpadlenc$ and $\qpadlftr$, and computationally-secure for $\qpadloop$ under the hardness of its underlying PRG and preprocessing model. 
\emph{(ii) PQ communication anonymity and unlinkability}, where the confidentiality of the anonymous transport reduces to the IND-CPA security of the layered encryption and the PQ security of the KEM, relying on standard symmetric-key assumptions together with Module-LWE-based hardness.  
\emph{(iii) Authenticity and event-scoped linkability of the proof artifact}, where the validity and misuse-detection properties of the proximity-backed proof artifact reduce to the simulation-extractability and soundness of the underlying SoK together with the one-wayness, collision resistance, and second-preimage resistance of the employed hash functions.  
\emph{(iv) Authentic and abuse-resistant service access}, where puzzle authenticity reduces to the EUF-CMA security of $\mldsa$ under standard module-lattice assumptions (e.g., Module-SIS/Module-LWE), while service-side computational throttling and replay/rate-limit enforcement reduce to the hardness of the selected puzzle construction and the hash-based binding of tokens, commitments, and event identifiers.
\end{theorem}

\vspace{-2mm}
\begin{mylemma}\label{lem:chor}
$\qpadlenc$ and $\qpadlftr$ achieve IT-secure private spectrum retrieval, where $\qpadlenc$ guarantees $(\ell-1)$-private query-index privacy and $\qpadlftr$ guarantees $t$-private query-index privacy together with $\nu$-Byzantine robustness and $k$-out-of-$\ell$ correctness under the Shamir-secret-sharing and decoding conditions.
\end{mylemma}

\vspace{-2mm}
\begin{mylemma}\label{lem:cippir}
$\qpadloop$ achieves computationally secure private spectrum retrieval against polynomial-time adversaries, where query-index privacy reduces to the pseudorandomness of the underlying PRG together with the preprocessing and chunk-permutation assumptions of the construction.
\end{mylemma}

\vspace{-2mm}
\begin{mylemma}\label{lem:hct}
$\qpadlhct$ provides hash-based computational throttling for service access, where the adversary’s success probability in bypassing the required client work reduces to the second-preimage resistance of the underlying hash function and the sequential structure of the $\hct$.
\end{mylemma}

\vspace{-2mm}
\begin{mylemma}\label{lem:lbp}
$\qpadllbp$ provides lattice-based computational throttling for service access, where the adversary’s success probability in bypassing the required client work reduces to the hardness of the underlying $\alpha$-Hermite Shortest Vector Problem instance family.
\end{mylemma}





\section{\textbf{Performance Evaluation and Comparison}} \label{sec:PerfEval} 
This section outlines our evaluation metrics and implementation setup, followed by a comprehensive assessment of $\qpadl$ across multiple instantiations, using diverse cryptographic techniques, optimizations, and GPU acceleration.

\subsection{Configuration and Experimental Setup} \label{subsec:Configuration} 
\noindent \textbf{Hardware:} We evaluated the efficiency of \qpadl~framework using a standard desktop equipped with an Intel Core $\textit{i9-11900K} @ 3.50~GHz$, $64~\text{GiB}$ RAM, $1~\text{TB}$ SSD, running Ubuntu $22.04.4~\text{LTS}$. It is also equipped with NVIDIA GTX 3060 GPU card, which provides CUDA 3584 cores, 12GB GDDR6-based memory, and 360GB/s memory bandwidth.

\smallskip
\noindent \textbf{Libraries:} We used \textit{C} and \textit{Python} programming languages along with several cryptographic libraries, including: \textit{percy++}\footnote{\url{https://percy.sourceforge.net/}} for multi-server PIR components, \textit{liboqs}\footnote{\url{https://openquantumsafe.org/}} for PQ-secure primitives, \textit{OpenSSL} for standard cryptographic operations such as hash functions, and \textit{NTL} for lattice-based puzzles. We have also used the LRS repository\footnote{\url{https://github.com/yuxi16/Post-Quantum-Linkable-Ring-Signature?tab=}} for the ring signature and the hashcash-tree repository\footnote{\url{https://github.com/alviano/hashcash-tree}} for the hash-based puzzle. Additionally, DBs were constructed using \textit{SQLite}\footnote{The DB is modeled as a matrix with adjustable row counts (e.g., $2^{10}$, $2^{12}$, $2^{14}$, $2^{16}$). For each grid cell ($l_x, l_y$), we generated synthetic spectrum records and embedded signed $\hct$ and $\lbp$ puzzles, storing them across $\psd$s synchronized per FCC requirements~\cite{grissa2021anonymous}.}. For CPU-bounded implementation, we consider AVX instructions and OpenMP \cite{chandra2001parallel} as in \cite{gunther2022gpu}.
\smallskip
\noindent {\textbf{Evaluation Metrics and Rationale:} 
We assess \qpadl~both analytically and empirically, measuring computational cost and communication overhead across all entities and spectrum access phases. Our assessment covers LRS costs in the $\pol$ phase, PIR overhead in $\qpadlenc$, $\qpadlftr$, and $\qpadloop$ under CPU and GPU implementations, $\psd$ DB setup costs, and PoW costs in $\qpadlhct$ and $\qpadllbp$, along with $\pqtor$ communication overhead. We further analyze scalability in terms of end-to-end delay under varying user loads, network conditions, and $\psd$ configurations. Since \qpadl~is the first framework to jointly achieve location privacy, anonymity, location verification, and counter-DoS with PQ security, direct comparisons are not feasible; instead, we provide a comprehensive standalone evaluation and contrast specific metrics with related works offering partial overlap.

\subsection{Experimental Results} \label{subsec:Performance} 
The analytical and empirical cryptographic, computational, and communication overhead across each phase of $\qpadl$ is shown in Tables~\ref{tab:analytical_costs} and \ref{tab:empirical_costs} and elaborated below:

\begin{table*}
\centering
\small
\caption{Analytical computational costs and communication overhead of $\qpadl$.} 
\label{tab:analytical_costs}
\begin{tabular}{|c|c||c|c|}
\hline
\textbf{Phase} & \textbf{Entity} & \textbf{Computational Cost} & \textbf{Communication Cost} \\\hline\hline
\multirow{2}{*}{\makecell[c]{\textbf{PoL} \\ $\qpadl$}} & 
\footnotesize User &
\makecell[c]{$O(n_l) \cdot H'+\, O(\log\log(n_l)) \cdot H'$} & 
\multirow{2}{*}{$O(\operatorname{polylog}(\log(n_\ap)))$} \\\cline{2-3}
& \footnotesize AP & \makecell[c]{$O(n_l) \cdot H'+\, O(\log(n_l) \cdot \log\log(n_l)) \cdot H'$} & 
\\\hline\hline
\textbf{Spectrum Query} & & & \\\hline
\multirow{2}{*}{\shortstack{\texttt{Puzzle.Bind} \\ ($\kappa=20$)}}
& \multirow{2}{*}{\footnotesize \makecell[c]{User \\ PSD}} &
\multirow{2}{*}{\makecell[c]{$16r \cdot O(n_\sigma \log n_\sigma + 4n_\sigma)+\, r \cdot n_\Lambda^3 \cdot \operatorname{mult}(n_\Lambda)$}} & 
\multirow{2}{*}{$r \cdot |\Pi|$} \\
& & & 
\\\hline
\multirow{1}{*}{$\qpadlenc$}
& \footnotesize \makecell[c]{User \\ PSD} &
\multirow{1}{*}{\makecell[c]{$(r + b) \cdot ((\ell - 1)\cdot t_\oplus)+ n \cdot t_\oplus$}} & 
$(r + b) \cdot \ell+O(\operatorname{polylog}(\log(n_l)))$ 
\\ \hline
\multirow{1}{*}{$\qpadlftr$}
& \footnotesize \makecell[c]{User \\ PSD} &
\multirow{1}{*}{\makecell[c]{$\ell(\ell{-}1)r t_\oplus +\, 3\ell(\ell{+}1)t_\oplus+(n/w)\cdot t_\oplus$}} & 
$r \cdot w \cdot \ell + k \cdot b+O(\operatorname{polylog}(\log(n_l)))$ 
\\ \hline
\multirow{1}{*}{$\qpadloop$}
& \footnotesize \makecell[c]{User \\ PSD} &
\multirow{1}{*}{\makecell[c]{$\sqrt{n} \cdot (r \cdot \ell + 1 + 1/\ell) \cdot t_\oplus +(n/2\ell)(1 + r - 1)\cdot t_\oplus$}} & 
$\ell \left(2\sqrt{n/(8\ell)} + \kappa/8\right) +O(\operatorname{polylog}(\log(n_l)))$
\\ \hline\hline
\textbf{Service Request} & & & \\\hline
\multirow{3}{*}{\makecell[c]{$\qpadlhct$ \\ ($n_l=2$)}}
& \footnotesize User &
$\lceil \log_2(n_l) \rceil \cdot 2^\kappa$ 
& \multirow{3}{*}{\makecell[c]{$\lambda + 32 +\lceil \log_2(n_l) \rceil \cdot |H|$ \\ $+O(\operatorname{polylog}(\log(n_l)))$}} \\\cdashline{2-3}
& \multirow{2}{*}{\footnotesize Server} & $\lceil \log_2(n_l) \rceil \cdot H+O(n_\sigma \log n_\sigma + 4n_\sigma)$ & \\
& & $+O(\log(n_l) \cdot \log\log(n_l)) \cdot H'$ &
\\\hline
\multirow{3}{*}{$\qpadllbp$}
& \footnotesize User &
$O(2^{0.2925 n_\Lambda + o(n_\Lambda)})$ & \multirow{3}{*}{\makecell[c]{$10n_\Lambda^2 + n_\Lambda(n_\Lambda{-}1)+10n_\Lambda^2+ n_\Lambda \log_2(\|\nu\|)$ \\ $+O(\operatorname{polylog}(\log(n_l)))$}} \\\cdashline{2-3}
& \multirow{2}{*}{\footnotesize Server} & $O(n_\Lambda^2 \cdot \operatorname{mult}(n_\Lambda))+O(n_\sigma \log n_\sigma + 4n_\sigma)$ & \\
& & $+O(\log(n_l) \cdot \log\log(n_l)) \cdot H'$ & \\\hline
\end{tabular}
\begin{tablenotes}[flushleft]
    \footnotesize
	\item Here, $\ell$ is the number of responsive $\psd$s, $w$ the number of words in the DB, $b$ the size of each DB item in bits, $r$ the number of rows, and $n = r \times b$ the total DB size in bits. $n_l$ denotes the number of leaves of $\hct$, $n_\ap$ the number of $\ap$s in the region, $n_\Lambda$ the lattice dimension, and $n_\sigma$ the lattice dimension in ML-DSA. $||\nu||$ represents the Euclidean norm of a lattice vector. $t_\oplus$ is the cost of an XOR operation, $H$ denotes the cost of a hash operation, and $H'$ the cost of the Rescue-Prime hash function. $mult(n_\Lambda)$ refers to multiplying two $n_\Lambda$-bit numbers. Finally, $\kappa$ indicates the puzzle difficulty level in the quantum setting. 
\end{tablenotes}\vspace{-4mm}
\end{table*}

\begin{table*}
\centering
\Large
\caption{Empirical computational performance of $\qpadl$.} 
\label{tab:empirical_costs}
\resizebox{\textwidth}{!}{
\begin{tabular}{|c||c||c|c|c|c|c|c|c|c|c|c|c|c|}
\hline
\textbf{Phase} & \textbf{Entity} & \multicolumn{12}{c|}{\textbf{Empirical Computational Cost (ms)}} \\\hline\hline

\textbf{PoL} & User & \multicolumn{12}{c|}{33.34} \\\cline{2-14}
$\qpadl$ & AP & \multicolumn{12}{c|}{49.37} \\\hline\hline

\multirow{2}{*}{\textbf{Spectrum Query}} & & 
\multicolumn{3}{c|}{\textbf{$|\db|=2^{12}$}} & 
\multicolumn{3}{c|}{\textbf{$|\db|=2^{14}$}} & 
\multicolumn{3}{c|}{\textbf{$|\db|=2^{16}$}} & 
\multicolumn{3}{c|}{\textbf{$|\db|=2^{18}$}} \\
& & 
$\qlen\boldsymbol{=1}$ & $\qlen\boldsymbol{=2^7}$ & $\qlen\boldsymbol{=2^{10}}$ & 
$\qlen\boldsymbol{=1}$ & $\qlen\boldsymbol{=2^7}$ & $\qlen\boldsymbol{=2^{10}}$ &
$\qlen\boldsymbol{=1}$ & $\qlen\boldsymbol{=2^7}$ & $\qlen\boldsymbol{=2^{10}}$ &
$\qlen\boldsymbol{=1}$ & $\qlen\boldsymbol{=2^7}$ & $\qlen\boldsymbol{=2^{10}}$ \\\hline

\multirow{4}{*}{\shortstack{\texttt{Puzzle.Bind} \\ ($\kappa=20$)}}
& \texttt{PSD}-$\lbp$ (CPU) & 
\multicolumn{3}{c|}{11816} & \multicolumn{3}{c|}{23320} & \multicolumn{3}{c|}{69820} & \multicolumn{3}{c|}{243602} \\
& \texttt{PSD}-$\lbp$ (GPU) & \multicolumn{3}{c|}{430} & \multicolumn{3}{c|}{1470} & \multicolumn{3}{c|}{5590} & \multicolumn{3}{c|}{22248.2} \\
& \texttt{PSD}-$\hcp$ (CPU) & \multicolumn{3}{c|}{344} & \multicolumn{3}{c|}{1376} & \multicolumn{3}{c|}{5500} &  \multicolumn{3}{c|}{22625.61} \\
& \texttt{PSD}-$\hcp$ (GPU) & \multicolumn{3}{c|}{346} & \multicolumn{3}{c|}{1370} & \multicolumn{3}{c|}{5470} &  \multicolumn{3}{c|}{21825.33} \\\hline

\multirow{3}{*}{$\qpadlenc$}
& User & \multicolumn{3}{c|}{0.258} & \multicolumn{3}{c|}{0.260} & \multicolumn{3}{c|}{0.273} & \multicolumn{3}{c|}{0.294} \\
 & \texttt{PSD} (CPU) & 8.99 & 11.54 & 30.03 & 10.46 & 33.59 & 245.93 & 17.24 & 176.88 & 1334.0 & 42.86 & 681.42 & 5309.90 \\
 & \texttt{PSD} (GPU) & 8.82 & 9.82 & 22.89 & 9.11 & 16.63 & 81.59 & 10.38 & 45.21 & 323.85 & 15.42 & 154.74 & 1269.30 \\ \hline

\multirow{3}{*}{$\qpadlftr$}
& User & \multicolumn{3}{c|}{3.62} & \multicolumn{3}{c|}{6.31} & \multicolumn{3}{c|}{7.51} & \multicolumn{3}{c|}{8.65} \\
 & \texttt{PSD} (CPU) & 38.9 & 170.50 & 1058.9 & 133.7 & 821.3 & 5435.85 & 507.8 & 3087.5 & 19878.41 & 4106.94 & 16785.87 & 111319.09 \\
 & \texttt{PSD} (GPU) & 22.18 & 37.60 & 149.79 & 55.41 & 102.77 & 574.64 & 199.22 & 389.76 & 2144.63 & 840.70 & 1720.35 & 9672.28 \\ \hline

\multirow{3}{*}{$\qpadloop$}
& User & \multicolumn{3}{c|}{2.02} & \multicolumn{3}{c|}{2.72} & \multicolumn{3}{c|}{5.30} & \multicolumn{3}{c|}{9.43} \\
 & \texttt{PSD} (CPU) & 8.95 & 10.52 & 23.84 & 9.08 & 13.68 & 42.81 & 9.93 & 21.01 & 113.41 & 22.71 & 58.41 & 369.22 \\
 & \texttt{PSD} (GPU) & 8.88 & 9.91 & 21.40 & 8.98 & 11.39 & 35.14 & 9.28 & 23.80 & 110.84 & 13.22 & 30.63 & 218.12 \\ \hline
\hline

\textbf{Service Request} & & \multicolumn{3}{c|}{$\mathbf{\kappa=14}$} & \multicolumn{3}{c|}{$\mathbf{\kappa=18}$} & \multicolumn{3}{c|}{$\mathbf{\kappa=20}$} & \multicolumn{3}{c|}{$\mathbf{\kappa=23}$} \\\hline

\multirow{3}{*}{\makecell[c]{$\qpadlhct$ \\ ($n_l=2$)}}
& User & \multicolumn{3}{c|}{38.94} & \multicolumn{3}{c|}{66.36} & \multicolumn{3}{c|}{316.56} & \multicolumn{3}{c|}{6251.37} \\\cdashline{2-13}
& \multirow{2}{*}{Server} & \multicolumn{12}{c|}{\multirow{2}{*}{10.53}} \\
& & \multicolumn{12}{c|}{ }\\\hline

\multirow{2}{*}{$\qpadllbp$}
& User & \multicolumn{3}{c|}{133.08} & \multicolumn{3}{c|}{259.15}  & \multicolumn{3}{c|}{881.41} & \multicolumn{3}{c|}{2931.45} \\\cdashline{2-13}
& \multirow{2}{*}{Server} & \multicolumn{3}{c|}{\multirow{2}{*}{9.331}} & \multicolumn{3}{c|}{\multirow{2}{*}{10.622}} & \multicolumn{3}{c|}{\multirow{2}{*}{11.304}} & \multicolumn{3}{c|}{\multirow{2}{*}{11.478}} \\
& & \multicolumn{3}{c|}{}&\multicolumn{3}{c|}{} &\multicolumn{3}{c|}{} &\multicolumn{3}{c|}{} \\\hline
\end{tabular}
}
\begin{tablenotes}[flushleft]
    \footnotesize
	\item All computation costs are reported in $ms$. We set classical security at 128 bits per NIST guidelines and PQ security to NIST Level I, equivalent to 128-bit classical strength~\cite{bavdekar2023post}, with all parameters aligned accordingly. $\hct$ uses SHA-256, and the LRS employs the Rescue-Prime hash function~\cite{szepieniec2020rescue} with a SoK based on ethSTARK~\cite{team2021ethstark}. 
\end{tablenotes}
\end{table*}

\subsubsection{Computational Costs} 
{\em \ul{(i) $\pol$ Phase}:} On the user side, this phase involves generating a location commitment via hashing and verifying the LRS, while the AP performs LRS signing. Signing includes an offline Merkle tree construction and root computation, followed by an online phase with Merkle path computation, hashing, and statement authentication using a SoK signature. Verification mirrors the offline setup and adds SoK verification. AP signing scales efficiently with ring size, from $20$~ms for $2^3$ members to $60$~ms for $2^{13}$, making it suitable for large SASs. Verification is lightweight, $0.4$~ms for $2^3$ users and $8$~ms for $2^6$, and $\ProxVerif$ adds only $1$–$10$~ms using signal strength and RTT.  
{\em \ul{(ii) Puzzle Binding and Database Setup}:} This phase involves generating $\hct$ and $\lbp$ puzzles, signing them with ML-DSA, and binding them to each DB entry across varying DB sizes. ML-DSA key generation, signing, and verification take approximately $29~\mu s$, $84~\mu s$, and $30~\mu s$, respectively. Puzzle generation for $\hct$ involves selecting a random string, while $\lbp$ requires generating a lattice basis using uniformly random numbers. The combined overhead of puzzle creation and signature operations across different DB sizes via CPU and GPU-parallelized implementation is detailed in Table~\ref{tab:empirical_costs}. 
{\em \ul{(iii) Spectrum Query Phase}:} This phase integrates three PIR schemes with $\pir.\query$, $\pir.\query.\response$, and $\pir.\blockrec$ operations, along with LRS verification and linkability checks on the $\psd$ side, all executed over $\pqtor$. PQ-Tor overhead includes circuit setup and layered encryption, primarily driven by three ML-KEM and AES operations. Specifically, ML-KEM key generation, encapsulation, and decapsulation take $10~\mu s$, $13.4~\mu s$, and $9~\mu s$, respectively, while AES-256 requires $7~\mu s$ for key generation and $8~\mu s$ for encryption. 
{\em \ul{(iv) Service Request Phase}:} This phase requires PoW on the client side using $\hct$ or $\lbp$, and on the server side, LRS verification and puzzle authentication via ML-DSA. Solving $\hct$ requires approximately $\lceil \log_2(n_l) \rceil \cdot 2^\kappa$ hash operations, taking 38-316 ms for $\kappa$ up to 18. In contrast, $\lbp$ employs lattice basis reduction and enumeration~\cite{albrecht2019general}, with runtimes ranging from 133-881 ms. For larger $\kappa$, lattice-based puzzles help maintain practical solving times (3 instead of 6 s) over hash-based ones while mitigating DoS attacks.

\subsubsection{Parallelization Assessment with GPU} 
Figure~\ref{fig:gpubench} shows the $\psd$ side runtime of $\qpadlenc$, $\qpadlftr$, and $\qpadloop$ for varying DB sizes and query counts ($\qlen$), with each DB entry fixed at 3KB. GPU-accelerated implementations achieve up to an order-of-magnitude speedup over CPU-bound versions. Among them, $\qpadloop$ has the lowest online server cost due to its offline-online design; for instance, with a 128MB DB and $2^{10}$ queries, it outperforms $\qpadlenc$ and $\qpadlftr$ by $3.2\times$ and $16.5\times$, respectively. This benefit, however, comes with $\approx19\times$ higher client computation, extra offline precomputation at each $\psd$, and an additional offline interaction with the user.  We emphasize that GPU performance gains over CPU increase with larger database and query sizes, particularly in $\qpadloop$, which more accurately reflects real-world requirements. Additionally, GPU parallelization can enhance the performance of \texttt{Puzzle.Bind}. For example, GPU-based construction of databases containing $\lbp$ provides up to $10\times$ speedup on databases with $2^{18}$ entries. For $\hct$, CPUs benefit from highly optimized OpenSSL hash functions, outperforming a single GPU core. Nevertheless, the GPU's advantage over the CPU becomes more pronounced as the database size exceeds $512$ MB. In GPU-accelerated PIR, keeping the static DB in GPU global memory avoids repeated CPU–GPU transfers, improving throughput. Our benchmarks demonstrate the advancement of GPU over CPU for multi-server PIRs and puzzle generation in \texttt{Puzzle.Bind} are available on GitHub \footnote{\url{https://github.com/kiarashsedghigh/GPU-Multiserver-PIR-PuzzleGen.git}}.

\begin{figure}
    \includegraphics[width=\linewidth]{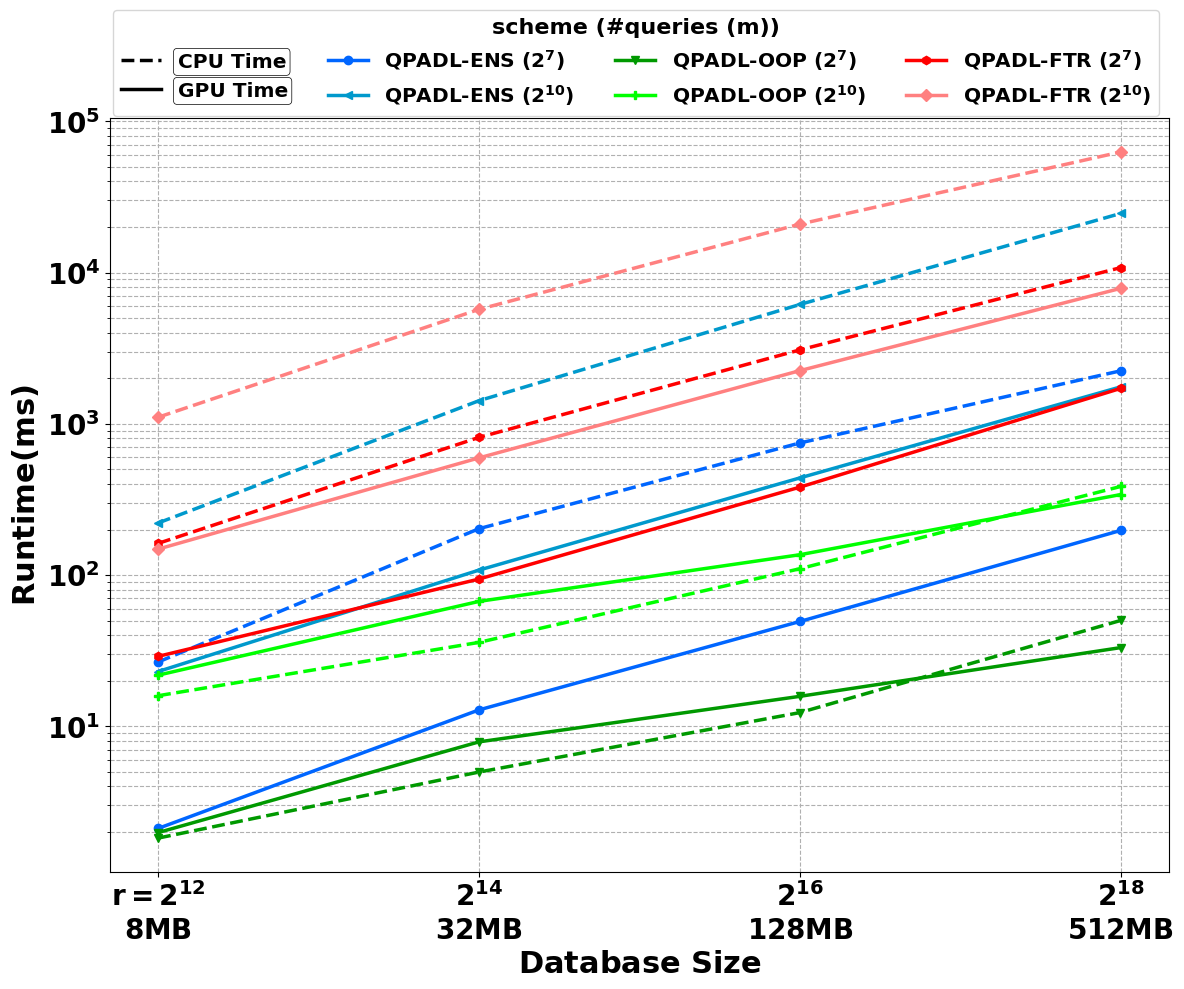}
    \caption{Scalability Benchmarking of PIRs on CPU/GPU}
    \label{fig:gpubench}
\end{figure}

\subsubsection{Communication Overhead} 
{\em (i)} The location commitment includes the location coordinates ($16$ bytes), beacon ($8$ bytes), timestamp ($8$ bytes), and a random nonce ($4$ bytes). The ring signature size scales with the number of ring members, measuring approximately $18$~KB for $2^3$ users and $20$~KB for rings of size $2^6$ to $2^{13}$. 

{\em (ii)} In $\qpadlenc$, $\qpadlftr$, and $\qpadloop$, the communication complexity involves retrieving a $b$-bit block from $\ell$ responsive $\psd$s, along with transmitting the location commitment and $\pol$. Each $b$-bit block contains $560$~bytes of spectrum data, an $\hct$ or $\lbp$ puzzle, and the $\psd$’s ML-DSA signature. The $\hct$ puzzle includes a $\lambda$-bit nonce ($n_s$), a $4$-byte difficulty ($\kappa$), and a $1$-byte level ($n_l$), totaling $37$~bytes. The $\lbp$ puzzle contains a $10n$-bit prime ($p$), $(n-1)$ samples of size $10n$ bits, plus $4$-byte values for $\alpha$ and $n_\Lambda$, totaling $28133$~bytes.    
Since spectrum queries run over PQ-Tor, its communication overhead directly impacts delays. PQ-Tor’s performance closely matches conventional Tor, with only minor differences from ML-KEM and AES-256 operations. Thus, we utilized conventional Tor network metrics for communication delay estimation~\cite{TorMetrics}. Although ML-KEM is faster than RSA (used in Tor), its use still requires two packet transmissions due to Tor’s $512$-byte packet size, resulting in an average circuit build time of $300$~ms. As each retrieved block is under $50$~KB, the PQ-Tor communication delay is bounded at approximately $175$~ms. 

{\em (iii)} In service request phase, the client transmits the location commitment, $\pol$, and the signed token. The $\hct$ solution size is $\lceil \log_2(n_l) \rceil \times |H|$ bits, while the $\lbp$ solution size is $n_\Lambda \times \lceil \log_2(2\alpha \cdot p^{1/n_\Lambda}) \rceil$ bits. The accompanying ML-DSA signature is $2420$~bytes.

\subsubsection{Scalability Assessment} 
We evaluate scalability through end-to-end (E2E) cryptographic delay, covering both computation and communication in spectrum access, including $\pol$ acquisition, spectrum query, and puzzle retrieval. Figure~\ref{fig:end2enddelay} shows communication domination for a few SUs, whereas with more SUs and larger DB sizes, computation becomes the bottleneck. GPU acceleration overcomes this: $\qpadlenc$ achieves $2.66$–$4.18\times$, $\qpadlftr$ $4.83$–$11.49\times$, and $\qpadloop$ $1.66$–$1.71\times$ speedup for up to $2^{10}$ SUs per grid and time window. Although GPU-accelerated $\qpadloop$ achieves the best performance, its speedup is smaller due to lower CPU overhead, while the gains for $\qpadlenc$ and $\qpadlftr$ grow with larger query volumes and DB sizes.

\begin{figure}
    \centering
    \includegraphics[width=\linewidth]{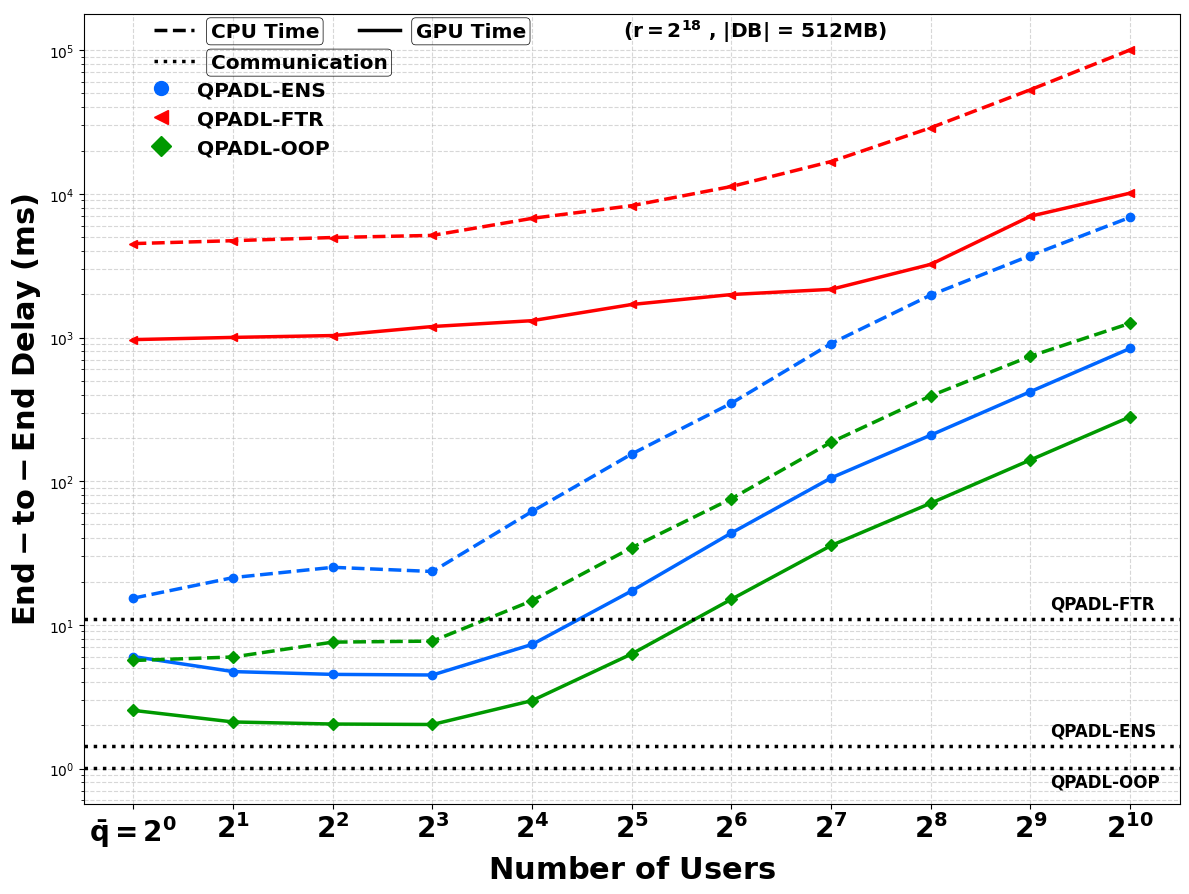}
    \caption{End-to-End Cryptographic Delay}
    \label{fig:end2enddelay}
\end{figure}

\subsubsection{Comparison with SOTA} 

Existing PIR-based schemes, such as~\cite{grissa2021anonymous, xin2016privacy, darzi2024privacy}, mainly protect the queried index, but they do not jointly provide strong communication anonymity, PQ security, and integrated service-side abuse resistance. From a practical perspective, $\qpadl$ also remains scalable: GPU acceleration yields $2.66$-$4.18\times$ speedup for $\qpadlenc$, $4.83$-$11.49\times$ for $\qpadlftr$, and $1.66$-$1.71\times$ for $\qpadloop$ for up to $2^{10}$ SUs per grid and time window, showing that the framework remains deployable even as the query load grows. Moreover, in the representative $\qpadloop$ instantiation, the PSD-side online cost remains as low as $110.84$~ms for $|DB|=2^{16}$ and $\bar{q}=2^{10}$ under GPU execution, compared with $323.85$~ms for $\qpadlenc$ and $2144.63$~ms for $\qpadlftr$ under the same setting. 

From the location-assurance perspective, only a limited subset of prior works incorporates explicit PoL or location-verification support, and these typically rely on trusted auxiliary entities, classical assumptions, or significantly higher latency. For example, Li et al.~\cite{li2015privacy} and Xin et al.~\cite{xin2016privacy} report PoL costs of about $210$~ms and $430.1$~ms, respectively, whereas $\qpadl$ achieves PQ-secure proximity-backed proof generation with only $33.34$~ms on the client side and $49.37$~ms on the AP side. This yields a combined cryptographic PoL cost of about $82.71$~ms, while additionally providing event-scoped misuse detection through the LRS-based proof artifact. Hence, unlike earlier query-private SAS schemes that largely assume honest users, $\qpadl$ directly accounts for replay, spoofing-oriented misuse, and proof forgery within the access workflow itself. 

From the communication and DoS-resilience perspectives, prior solutions usually optimize one dimension at the expense of the others. Existing privacy-preserving schemes incur substantial communication overhead, e.g., about $753$~KB for LP-Chor~\cite{grissa2019location}, $6$~MB for LP-Goldberg~\cite{grissa2019location}, $125$~KB for RAID-LP-Chor~\cite{grissa2019location}, $1.25$~MB for TrustSAS~\cite{grissa2021anonymous}, and $605.92$~KB for PACDoSQ~\cite{darzi2024privacy}, while Xin et al.~\cite{xin2016privacy} still requires $325$~KB in a single-DB setting. In contrast, each retrieved block in $\qpadl$ remains below $50$~KB, and the PQ-Tor communication delay is bounded at about $175$~ms, with an average circuit-build time of about $300$~ms. At the service layer, $\qpadl$ further embeds authenticated computational throttling directly into the workflow: HCT solving costs range from $38.94$~ms to $316.56$~ms for moderate difficulty levels, while the lattice-based alternative provides PQ-secure throttling with client-side costs from $133.08$~ms to $881.41$~ms. Therefore, unlike detection-only defenses or isolated request-throttling methods, $\qpadl$ offers a more complete and deployment-oriented security architecture for SAS operation during the PQ transition.

\section{Conclusion and Future Work} \label{sec:Conclusion} \vspace{-1mm}
This work introduced $\qpadl$, the first framework to simultaneously address location privacy, anonymity, location spoofing, and DoS threats in SASs under PQ security assumptions. By integrating privacy-preserving spectrum queries with robust CPPs, $\qpadl$ mitigates both conventional and PQ threats while maintaining scalability for large-scale SAS. Our proposed instantiations, built on diverse cryptographic primitives, offer flexible security-performance trade-offs. Formal security analysis confirms the framework’s resilience, and extensive performance evaluations, enhanced through GPU parallelization and optimization, demonstrate its practicality and efficiency, establishing $\qpadl$ as a viable and future-proof solution for secure spectrum access. 

\appendix
\section{APPENDIX} \label{sec:appendix_security}

\subsection{Security Analysis}
\label{sec:SecuirtyAppendix} 

We give a series of security proofs capturing the threat and security models as follows:  \vspace{+1mm}

\noindent \textbf{Theorem 1.} \textit{For any QPT adversary interacting with $\qpadl$, the framework achieves, except with probability $negl(\lambda)$: \emph{(i) PQ-secure location private spectrum access}, where the privacy of the queried spectrum index reduces to the security of the instantiated PIR layer, namely IT-secure for $\qpadlenc$ and $\qpadlftr$, and computationally-secure for $\qpadloop$ under the hardness of its underlying PRG and preprocessing model. 
\emph{(ii) PQ communication anonymity and unlinkability}, where the confidentiality of the anonymous transport reduces to the IND-CPA security of the layered encryption and the PQ security of the KEM, relying on standard symmetric-key assumptions together with Module-LWE-based hardness.  
\emph{(iii) Authenticity and event-scoped linkability of the proof artifact}, where the validity and misuse-detection properties of the proximity-backed proof artifact reduce to the simulation-extractability and soundness of the underlying SoK together with the one-wayness, collision resistance, and second-preimage resistance of the employed hash functions.  
\emph{(iv) Authentic and abuse-resistant service access}, where puzzle authenticity reduces to the EUF-CMA security of $\mldsa$ under standard module-lattice assumptions (e.g., Module-SIS/Module-LWE), while service-side computational throttling and replay/rate-limit enforcement reduce to the hardness of the selected puzzle construction and the hash-based binding of tokens, commitments, and event identifiers.}

\begin{proof} 
{\em (i)} \textit{\ul{$t$-private PQ-secure Location Privacy}:} $\qpadl$ is instantiated with PIR schemes that ensure $t$-private location privacy during spectrum access, meaning that any coalition of at most $t$ corrupted $\psd$s learns nothing about the queried index $\theta$ beyond negligible probability, as formalized in Definition~\ref{def:tprivate}. In particular, both $\qpadlenc$ and $\qpadlftr$ provide information-theoretic privacy: $\qpadlenc$ achieves privacy against up to $\ell-1$ colluding servers, while $\qpadlftr$ achieves $t$-private retrieval together with its robustness guarantees. Hence, for any QPT adversary $\mathcal{A}$ corrupting a subset $S \subseteq [\ell]$ with $|S|\le t$ and for any two query indices $\theta_0,\theta_1 \in \db$, $\left|
\Pr\!\left[\mathcal{A}\!\left(\mathsf{View}^{\texttt{PIR}}_{S}(\theta_0)\right)=1\right] -\Pr\!\left[\mathcal{A}\!\left(\mathsf{View}^{\texttt{PIR}}_{S}(\theta_1)\right)=1\right]
\right|
\leq negl(\lambda)$. For the information-theoretic instantiations, this bound is independent of the adversary’s computational power, including quantum capabilities. In contrast, the $\qpadloop$ provides computational PIR privacy, where any non-negligible distinguishing advantage against the queried index yields an adversary against the hardness of the underlying PRG and preprocessing model, as formalized in Lemmas~\ref{lem:chor} and~\ref{lem:cippir}.


{\em (ii)} \textit{\ul{PQ Communication Anonymity and Unlinkability}:}
$\qpadl$ preserves client anonymity and unlinkability against $\psd$s and external observers through $\pqtor$ with three relays $(N_e,N_m,N_x)$, where each relay learns only its predecessor and successor, as formalized in Definition~\ref{def:anonymity}. Messages are transmitted over an onion circuit with layered symmetric encryption, $ctxt=\texttt{Enc}_{\sk_{N_e}}\!\big(\texttt{Enc}_{\sk_{N_m}}(\texttt{Enc}_{\sk_{N_x}}$ $(m))\big)$, where the session keys $\sk_{N_i}\in\{0,1\}^{256}$ are established via PQ-secure KEM encapsulation and decapsulation, $ctxt'_i \gets \texttt{ML}\mbox{-}\texttt{KEM}.\texttt{Encaps}(\pk_{N_i,\kyber}),  \sk_{N_i}\gets \texttt{ML}\mbox{-}\texttt{KEM}.\texttt{Decaps}$ $(\sk_{N_i,\kyber},ctxt'_i)$, for $i\in\{e,m,x\}$. Hence, for any QPT adversary $\mathcal{A}$ attempting to distinguish between two executions that differ only in the honest sender or receiver, its advantage is bounded by the IND-CPA security of the layered encryption and the Module-LWE-based hardness underlying the KEM, i.e., $Adv_{\mathcal{A}}^{\pqtor\text{-}\mathsf{Anon}}(\lambda)
\le
Adv_{\mathcal{B}_1}^{\mathsf{IND\mbox{-}CPA}}(\lambda)
+
Adv_{\mathcal{B}_2}^{\mathsf{MLWE}}(\lambda)
+negl(\lambda)$. Thus, any non-negligible break of communication anonymity or unlinkability in $\qpadl$ yields an adversary against either the confidentiality of the layered encryption or the PQ security of the underlying KEM. Concretely, this relies on AES-256 offering about 128-bit post-quantum security under Grover’s algorithm~\cite{bonnetain2019quantum}, together with the security of Module-LWE-based KEMs, whose hardness is related to worst-case module-lattice problems such as MSIVP in the random oracle model~\cite{bos2018crystals}. 


{\em (iii)} \textit{\ul{Proximity-Backed Location Assurance and Spoofing Resistance}:} In $\qpadl$, proximity-backed location assurance relies on the authenticity and event-scoped linkability of its PQ LRS scheme~\cite{xue2024efficient}, built in the ROM. The scheme is instantiated via a hash-based non-interactive argument of knowledge (NIAoK), namely an augmented zero-knowledge variant of ethSTARK transformed into a SoK using the Fiat--Shamir heuristic. The signer’s $\sk_l$ is committed as $\pk_l := H'(\sk_l)$, and a coalition of user keys is embedded in a Merkle tree to define the ring $\mathsf{R} = \{\pk_1, \dots, \pk_n\}$, where $n = 2^k$ and $k = \lceil \log_2(n) \rceil$. The relation proven in the SoK is formalized as $\mathsf{R}_s = \{((e, rt, T),(P,l,\sk_l)) : T = H'(\sk_l,e),\; rt = \texttt{MPath}(H'(\sk_l),\mathbf{P},l)\}$, where $e \in \mathbb{F}_p$ is the event identifier, $T$ is the linkability tag, $rt$ is the Merkle root, $l$ is the binary index of $\pk_l$ in the tree, and $\mathbf{P}$ is the corresponding Merkle path. For any two signatures $\sigma_1,\sigma_2$ on messages $m_1,m_2$, event-scoped linkability holds whenever their tags satisfy $H'(\sk_l,e_1)=H'(\sk_l,e_2)$. Thus, reuse of a proof artifact under the same event scope is detectable. The zero-knowledge property of the NIAoK ensures that no information beyond validity leaks, while unforgeability reduces to the collision resistance and preimage resistance of $H'$, together with the non-slanderability and extractability of the underlying SoK (\cite{xue2024efficient} for details). Under the structured-hash assumptions and the one-wayness of $\texttt{MPath}$, any adversary $\mathcal{A}$ attempting to forge a valid proof artifact or violate event-scoped linkability has success probability bounded by $Adv^{\text{forge}}_{\mathcal{A}}(\lambda)
\leq
Adv^{\text{SoK}}_{\mathcal{A}'}(\lambda)
+
Adv^{\text{Hash}}_{\mathcal{A}''}(\lambda)
\leq
negl(\lambda)$, where $\mathcal{A}'$ breaks SoK extractability and $\mathcal{A}''$ finds a hash collision or preimage. Therefore, within the assumed proximity-verification model of $\qpadl$, the resulting proof artifact is authentic, unlinkability-preserving across different event scopes, and misuse-detectable within the same event scope, thereby providing cryptographic protection against proof forgery, replay, and related spoofing attempts without claiming full physical-layer secure localization (cf. Definition~\ref{def:locationverification}).

{\em (iv)} \textit{\underline{Authentic and Abuse-Resistant Service Access}:} In \qpadl, puzzle authenticity is enforced through the existential unforgeability of $\mldsa$ under chosen-message attacks (EUF-CMA). Let $(\Pi_\theta,\sigma_\theta)$ denote a valid puzzle-signature pair. For any PPT adversary $\mathcal{A}$, the advantage of forging a valid pair not issued by an authorized $\psd$ is bounded by $Adv_\mathcal{A}^{\mathsf{EUF\mbox{-}CMA}}(\lambda)
:=
\Pr[(\Pi_\theta^\ast,\sigma_\theta^\ast)\leftarrow \mathcal{A} \text{ and } (\Pi_\theta^\ast,\sigma_\theta^\ast)\notin \mathcal{Q}]
\leq negl(\lambda)$, where $\mathcal{Q}$ is the set of honestly issued puzzle-signature pairs. This guarantee relies on the standard module-lattice assumptions underlying $\mldsa$, namely Module-LWE and Module-SIS hardness, with security reducing tightly to MSIS in the quantum ROM; hence, only puzzles issued by an authorized $\psd$ are accepted except with negligible probability.  

The service-side anti-abuse mechanism in $\qpadl$ is instantiated through CPPs, relying either on the second-preimage resistance of the underlying hash function in $\qpadlhct$ or on the hardness of the Hermite-SVP problem in $\qpadllbp$, as formalized in Lemmas~\ref{lem:hct} and~\ref{lem:lbp}, respectively. In addition, per-client rate-limiting at $\psd$s relies on the event-scoped binding of commitments and the linkability of the proof artifact. Specifically, the committed value is of the form $
C_\tw = H((l_x,l_y)\,\|\,\beta_\tw\,\|\,\tw\,\|\,r)$, where the location, beacon, active time window, and nonce are cryptographically bound by the one-wayness, collision resistance, and second-preimage resistance of $H$. An adversary $\mathcal{A}$ attempting to bypass replay or one-use-per-window enforcement must either produce a fresh valid proof artifact for the same event scope without the required witness, break event-scoped linkability, or find distinct openings that yield the same commitment, i.e., $H((l_x,l_y)\,\|\,\beta_\tw\,\|\,\tw\,\|\,r)
=
H((l_x',l_y')\,\|\,\beta_\tw'\,\|\,\tw'\,\|\,r')$, for two distinct tuples. Thus, any successful replay, token misuse, or rate-limit bypass implies either a break of the proof-artifact assumptions in part~(iii), a violation of the binding properties of $H$, or a break of the selected CPP hardness assumption. Therefore, $\qpadl$ provides authenticated puzzle issuance, per-request computational throttling, and abuse-resistant service access under the stated assumptions.
\end{proof}

\noindent\textbf{Lemma 1.} \textit{$\qpadlenc$ and $\qpadlftr$ achieve IT-secure private spectrum retrieval, where $\qpadlenc$ guarantees $(\ell-1)$-private query-index privacy and $\qpadlftr$ guarantees $t$-private query-index privacy together with $\nu$-Byzantine robustness and $k$-out-of-$\ell$ correctness under the Shamir-secret-sharing and decoding conditions.}

\begin{proof} 
In $\qpadlenc$, the client selects $r$-bit binary strings ${\{\rho_i\}_{i=1}^{\ell-1} \in \mathbb{GF}(2)^r}$ uniformly at random and sets $\rho_\ell := \bigoplus_{i=1}^{\ell-1} \rho_i \oplus e_\theta$, with $e_\theta$ being the unit vector at position $\theta$. The final response is $D_\theta := \bigoplus_{i=1}^\ell \rho_i$. For any coalition of corrupted servers $C \subset \{1,\dots,\ell\}$ with $|C| \leq \ell{-}1$, and for any pair of indices $\theta', \theta'' \in [r]$, it holds that $\Pr[\{\rho_i\}_{i \in C} \mid \theta = \theta'] = \Pr[\{\rho_i\}_{i \in C} \mid \theta = \theta'']$, implying zero distinguishing advantage between queries. This shows that the distribution of queries received by each $\psd$ is independent of $\theta$. Assuming all servers respond honestly, $\qpadlenc$ ensures perfect (IT) privacy during the query phase.  In $\qpadlftr$, the client encodes $\theta$ using $r$ random degree-$t$ polynomials $\{f_j(x)\}$ over $\mathbb{F}[x]$ such that $f_j(0) = e_\theta[j]$, and sends to the $\psd_j$ the query $\rho_j := \langle f_1(\alpha_j), \dots, f_r(\alpha_j) \rangle$, receiving response $R_j := \rho_j \cdot \db$. For any coalition $C \subset [\ell]$ with $|C| \leq t$, and any $\theta', \theta'' \in [r]$, it holds that $\Pr[\{\rho_i\}_{i \in C} \mid \theta = \theta'] = \Pr[\{\rho_i\}_{i \in C} \mid \theta = \theta'']$, yielding unconditional privacy for the target index $\theta$ and zero advantage for any unbounded $\mathcal{A}$. Moreover, $\qpadlftr$ ensures block reconstruction despite failures or malicious servers by using the Guruswami-Sudan list decoding algorithm, which corrects up to $\nu < k - \lfloor \sqrt{kt} \rfloor$ Byzantine responses when $k > t$ servers reply under $(\ell,t)$-Shamir secret sharing.
\end{proof}

\noindent\textbf{Lemma 2.} \textit{$\qpadloop$ achieves computationally secure private spectrum retrieval against polynomial-time adversaries, where query-index privacy reduces to the pseudorandomness of the underlying PRG together with the preprocessing and chunk-permutation assumptions of the construction.}

\begin{proof} 
Let the DB be identically replicated across $\ell$ servers, with each block split into $\pi$ chunks $\{x^{(j)}\}_{j=1}^{\pi}$, and with chunk order differing across servers. During preprocessing, SU selects a random $\lambda$-bit seed and applies a secure PRG to derive $\pi$ query vectors. Responses are computed as ${R_i := A_i \oplus q_i \cdot \texttt{chunk}_0}$, where ${A := q \cdot \db}$ and ${q := \texttt{PRG}(S_i, k(t{-}1))}$, as shown in Fig.~\ref{fig:instantiation}. Assuming a secure PRG, the $\qpadloop$ achieves PQ-secure computational $\pi$-private location privacy against any coalition of $C \subset [\ell]$ with $|C| < \ell$. Since the PRG outputs are indistinguishable from uniform and the flip chunk is hidden from $C$, the $\mathcal{A}$’s view ${\{{\rho_i}^{(j)}\}_{i \in C, j \in [\pi]}}$ is computationally indistinguishable for any pair ${\theta', \theta'' \in [r]}$, satisfying: $|\Pr[\mathcal{A}(\{{\rho_i}^{(j)}\}_{i \in C}) \mid \theta = \theta'] -\Pr[\mathcal{A}(\{{\rho_i}^{(j)}\}_{i \in C}) \mid \theta = \theta'']| \leq negl(\lambda)$, where the distinguishing advantage is bounded as: $Adv^{\text{Privacy}}_{\mathcal{A}}(\lambda) := \max_{\theta', \theta''}$ $\left| \Pr[\mathcal{A}(\textit{view}_{\theta'})] - \Pr[\mathcal{A}(\textit{view}_{\theta''})] \right|$ $\leq Adv^{\text{PRG}}_{\mathcal{A}'}(\lambda)$ with $\mathcal{A}'$ being a reduction that breaks the PRG’s PQ security. 
\end{proof}

\noindent \textbf{Lemma 3.} \textit{$\qpadlhct$ provides hash-based computational throttling for service access, where the adversary’s success probability in bypassing the required client work reduces to the second-preimage resistance of the underlying hash function and the sequential structure of the $\hct$.}

\begin{proof}
In $\qpadlhct$, the client solves a sequence of $\ell=\lceil \log_2(n_l)\rceil$ hashcash puzzles along a selected path in a binary hash tree with $n_l$ leaves, where each puzzle instance is parameterized by $\pi=(h,n_s,\kappa,n_l)$. For each node on the path, the probability that a $\mathcal{A}$ finds a valid solution in one trial is $2^{-\kappa}$, so the expected work per puzzle is $O(2^\kappa)$ classically and $O(2^{\kappa/2})$ under Grover-style quantum search. Since the verifier checks a valid root commitment together with all puzzle solutions along the selected path, the total expected work per request is $\ell \cdot 2^\kappa = \lceil \log_2(n_l)\rceil \cdot 2^\kappa$ in the classical setting, with a corresponding quadratic quantum speedup in the brute-force search term only.  
Moreover, because each higher-level node depends on the validity of its descendants, the path must be solved in sequence; thus, the construction enforces at least $\lceil \log_2(n_l)\rceil$ sequential proof-of-work steps and cannot be fully parallelized. Any adversary attempting to bypass this work must either solve all path puzzles successfully or produce a distinct invalid path or node assignment that still verifies against the committed root, which would violate the second-preimage resistance of the underlying hash function. Therefore, the adversary’s success probability is bounded by $
Adv_{\mathcal{A}}^{\mathsf{HCT}}(\lambda)
\leq
\ell \cdot 2^{-\kappa}
+
Adv_{\mathcal{B}}^{\mathsf{2nd\mbox{-}Pre}}(\lambda)$, where $\ell=\lceil \log_2(n_l)\rceil$. For standard parameter choices, this quantity is negligible in $\lambda$ (or in $\kappa$ when $\kappa$ is selected as a function of $\lambda$). Hence, $\qpadlhct$ provides verifiable hash-based throttling against request flooding under the second-preimage resistance of $\mathcal{H}$ and the sequential-work structure of the $\hct$.
\end{proof}


\noindent \textbf{Lemma 4.} \textit{$\qpadllbp$ provides lattice-based computational throttling for service access, where the adversary’s success probability in bypassing the required client work reduces to the hardness of the underlying $\alpha$-Hermite Shortest Vector Problem instance family.}

\begin{proof} 
In $\qpadllbp$, the lattice-based PoW instance $\pi = (\alpha, n_\Lambda, B, p)$ relies on the hardness of the $\alpha$-Hermite Shortest Vector Problem. Specifically, the challenge is to compute a nonzero vector of the lattice $\mathbf{v}\in \Lambda(B)\backslash\{0\}$ such that $||\mathbf{v}||\leq \alpha\cdot\lambda_1(\Lambda)$, where $\lambda_1(\Lambda) = \min\{||u|| : u \in \Lambda \backslash\{0\}\} \leq p^{1/n_\Lambda}$ is the length of the shortest nonzero lattice vector, and $\alpha$ is a tunable approximation factor set to $\alpha = 1.05 \cdot \Gamma(n_\Lambda/2+1)^{1/n_\Lambda} / \sqrt{\pi}$, with $n_\Lambda$ denoting the lattice dimension. Thus, producing a valid service token requires solving an $\alpha$-Hermite-SVP instance over $\Lambda(B)$. 
Using the best-known attacks, such as lattice reduction and enumeration, solving these puzzles requires on average $O(2^{0.2925 \times n_\Lambda + o(n_\Lambda)})$ operations in the classical setting and $O(2^{0.2570 \times n_\Lambda + o(n_\Lambda)})$ operations in the quantum setting. Hence, for any PPT or QPT adversary $\mathcal{A}$ attempting to obtain a valid token without performing the prescribed work, its success probability is bounded by the hardness of the underlying lattice problem, i.e., $Adv_{\mathcal{A}}^{\mathsf{LBP}}(\lambda)
\leq
Adv_{\mathcal{B}}^{\mathsf{Hermite\mbox{-}SVP}}(\lambda)
+negl(\lambda)$, where $\mathcal{B}$ is a reduction that uses $\mathcal{A}$ to solve the corresponding $\alpha$-Hermite-SVP instance. Equivalently, this implies $\Pr[\mathcal{A}\text{ finds }\mathbf{v}\in \Lambda(B)\text{ s.t. }||\mathbf{v}||\leq \alpha\cdot\lambda_1(\Lambda(B))]
\leq 2^{-\kappa}$,for security parameter $\kappa$ induced by the selected lattice parameters. Each puzzle remains valid only for a limited duration determined by $\kappa$, which in turn depends on $\alpha$, $\lambda$, and $n_\Lambda$. Therefore, except with negligible probability, $\qpadllbp$ enforces verifiable client work and provides abuse-resistant service access under the hardness of the selected lattice parameters.
\end{proof}


\printcredits

\section*{Acknowledgment} \label{sec:ack} 
This work is supported by the NSF-SNSF (2444615) and NSF CNS-2350213.


\bibliographystyle{cas-model2-names}  
\bibliography{SalehRef}             

\end{document}